\documentclass{aa}

\usepackage[T1]{fontenc}
\usepackage[varg]{txfonts}
\usepackage{mathptmx}
\usepackage{amsmath}
\usepackage{amssymb}
\usepackage{amstext}
\usepackage{amsfonts}
\usepackage{mathrsfs}
\usepackage{textcomp}

\usepackage{graphicx}

\bibpunct{(}{)}{;}{a}{}{,}
\defcitealias{Schafer2020}{SJB20}
\defcitealias{Schafer2022}{SJ22}
\defcitealias{Flock2020}{F20}

\newcommand{\dust}{\textrm{d}}
\newcommand{\gas}{\textrm{g}}
\newcommand{\mm}{\textrm{mm}}
\newcommand{\cm}{\textrm{cm}}
\newcommand{\mum}{\text{\textmu}\textrm{m}}
\newcommand{\au}{\textrm{au}}
\newcommand{\yr}{\textrm{yr}}
\newcommand{\kyr}{\textrm{kyr}}
\newcommand{\St}{\textrm{St}}
\newcommand{\K}{\textrm{K}}
\newcommand{\g}{\textrm{g}}

\begin{document}

\title{The coexistence of the streaming instability and the vertical shear instability in protoplanetary disks}
\subtitle{Scale-dependence of dust diffusion}
\author{Urs Sch\"afer\inst{\ref{Copenhagen}}
\and Anders Johansen{\inst{\ref{Copenhagen},\ref{Lund}}}
\and Mario Flock\inst{\ref{Heidelberg}}}
\institute{Centre for Star and Planet Formation, Globe Institute, University of Copenhagen, {\O}ster Voldgade 5-7, 1350 Copenhagen, Denmark, \email{urs.schafer@sund.ku.dk}\label{Copenhagen}
\and Lund Observatory, Department of Physics, Lund University, Box 43, 22100 Lund, Sweden\label{Lund}
\and Max-Planck-Institut f\"ur Astronomie, K\"onigsstuhl 17, 69117 Heidelberg, Germany\label{Heidelberg}}
\date{}
\abstract{The vertical shear instability and the streaming instability are two robust sources of turbulence in protoplanetary disks. The former has been found to induce anisotropic turbulence that is stronger in the vertical than in the radial dimension and to be overall stronger compared to the largely isotropic turbulence caused by the streaming instability. In this study, we shed light on the dust diffusion by the vertical shear instability and the streaming instability separately and together, and in particular on the direction- and scale-dependence of the diffusion. To this end, we employ two-dimensional global models of the two instabilities either in isolation or in combination. The vertical shear instability in isolation diffuses dust more strongly in the vertical direction than the streaming instability in isolation, resulting in a wave-shaped dust layer in our two-dimensional simulations. Compared with this large-scale diffusion, though, our study highlights that the vertical shear instability causes substantially weaker or even negligible small-scale diffusion. We validate this result using previously published three-dimensional simulations. In particular when simulating centimetre-sized dust, the undulating dust layer becomes internally razor-thin. In contrast, the diffusion owing to the streaming instability exhibits only a marginal scale-dependence, with the dust layer possessing a Gaussian shape. In models including both instabilities, the undulating mid-plane layer is broadened to a width set by the intrinsic diffusion level caused by the streaming instability.}
\titlerunning{Scale-dependence of dust diffusion}
\authorrunning{U. Sch{\"a}fer, M. Flock and A. Johansen}
\maketitle

\section{Introduction}
Observations reveal that protoplanetary disks are turbulent, with this turbulence generally being subsonic \citep[e.g.][]{Rosotti2023}. Up to supersonic turbulent speeds are only reached in the upper disk layers or very close to the central star \citep{Carr2004, Najita2009, Doppmann2011}. This renders turbulence measurements based on the broadening of molecular emission lines challenging since thermal and turbulent broadening are difficult to disentangle.

Nevertheless, owing in particular to the unprecedented spatial and spectral resolution of the Atacama Large Millimeter/submillimeter Array (ALMA), line broadening observations have yielded turbulent strengths in the outer regions of a handful of disks. In the disks surrounding HD163296, TW Hya, MWC480, and V4046 Sgr, these strengths have been constrained to be at most~\mbox{${\sim}1\%$} of the sound speed \citep{Flaherty2015, Flaherty2017, Flaherty2018, Flaherty2020, Teague2018}. Yet, in DM Tau and IM Lup disks Mach numbers of~$0.2-0.6$ have been inferred \citep{Flaherty2020, Rosotti2023, PanequeCarreno2023, Flaherty2024}. Moreover, ALMA observations have enabled not only measurements of the magnitude of gas turbulence, but also the detection of structures in the gas kinematics \citep[e.g.][]{Pinte2023}. Although structures like meridional flows \citep{Teague2019, Yu2021} have been interpreted as indications of the presence of planets, they could also be an imprint of the vertical shear instability \citep{BarrazaAlfaro2021, BarrazaAlfaro2024}. 

Further observational constraints on the strength of gas turbulence can be obtained indirectly from the vertical and radial distribution of dust. These are degenerate with the strength of the drag coupling between dust and gas, however. The dust scale height, on the one hand, constitutes a means to measure gas turbulence since it is regulated by the vertical stellar gravity and turbulent diffusion. From dust-to-gas scale height ratios of at most~$0.1$, \citet{Pinte2016} and \citet{Villenave2022} infer turbulent speeds of~\mbox{${\sim}1\%$} of the sound speed or less in the disks around HL Tau and Oph163131, respectively \citep[see also][]{Ueda2021}. Dust scale heights ranging between~$0.1$ and one gas scale height as well as Mach numbers as low as~\mbox{$\lesssim0.05$} and as high as~\mbox{$0.1-0.5$} are derived in different parts of the ring-and-gap structure of the HD163296 disk \citep{Ohashi2019, Doi2021, Liu2022} -- complementary with the estimates for this disk obtained from line broadening.

Similar to the scale height, the radial width of the dust rings observed in a number of disks has been utilised to measure gas turbulence \citep{Dullemond2018, Sierra2019, Rosotti2020, Facchini2020, Carvalho2024}. If the rings are associated with gas pressure maxima, turbulent diffusion counteracts the tendency of the dust to drift towards the centre of the rings. This approach has yielded a Mach number of~$0.1$ in the disk surrounding HD169142 \citep{Sierra2019} and values that are comparable or smaller by up to an order of magnitude in the disks around AS 209, HD163296, and Elias 2-24 \citep{Rosotti2020, Zagaria2023, Carvalho2024}. Differences in the turbulent strengths derived from the radial and vertical dust distributions might be indicative of anisotropic turbulence \citep{Doi2021, Villenave2022, Zagaria2023}.

Recent years have seen a rapid development of our understanding of protoplanetary disks turbulence not only from an observational perspective. Theoretically, a variety of instabilities have been established as potential sources and a map of the dominant mechanisms to drive turbulence at different disk radii and heights is emerging \citep[e.g.][]{Lesur2023}. Nevertheless, this picture is not yet comprehensive as we are only beginning to apprehend the intricacies of disk turbulence, including the interaction of different instabilities and how this interplay varies with length scale.  

In \citet[][hereafter SJB20]{Schafer2020} and \citet[][hereafter SJ22]{Schafer2022}, we explore the coexistence of two among the most promising instabilities, the vertical shear instability and the streaming instability. The former is a hydrodynamic instability that accesses the free energy associated with -- as the name indicates -- vertical shear \citep{Barker2015}. Vertical shear arises, for instance, from baroclinity, that is to say a misalignment between the density and pressure gradients. Owing to temperature gradients, protoplanetary disks are indeed generally baroclinic. Since it is generally dampened by vertical buoyancy, the instability requires gas cooling to be sufficiently rapid \citep{Nelson2013, Lin2015}. This condition is met in the outer disk regions at radii~\mbox{$\gtrsim10~\au$} \citep{Lin2015, Malygin2017, Pfeil2019}. We note that the linear analysis conducted by \citet{Latter2018} shows that the vertical shear instability experiences growth even at the longer cooling times relevant in optically thick regions, albeit with a reduced growth rate.

The vertical shear instability drives anisotropic turbulence, with the turbulent gas speeds being much larger in the vertical than in the radial dimension (\citealt{Nelson2013}, \citealt{Stoll2017}, \citetalias{Schafer2020}). In models including only the drag coupling of the dust to the gas (not vice versa), the vertical gas motions stir up dust to a scale height that is comparable to the gas scale height even if the Stokes number of the dust -- the ratio of the stopping time, the timescale of this drag coupling, to the dynamical timescale -- is as large as~$0.1$ \citep{Stoll2016, Flock2017, Dullemond2022}. However, if the drag that the dust exerts on the gas is taken into account, the tendency of the dust to sediment to the disk mid-plane ``weighs'' down the gas. This effective buoyancy causes a reduction of both the vertical turbulent gas velocities and the dust scale height (\citealt{Lin2017}, \citealt{Lin2019}, \citetalias{Schafer2020}), the latter to~\mbox{${\sim}10\%$} of the gas scale height for the above Stokes number and the canonical dust-to-gas surface density ratio of~$1\%$ \citep{Lin2019}.

The streaming instability, as originally discovered by \citet{Youdin2005}, arises where dust and a (negative) radial gas pressure gradient are present, that is all but everywhere in the dust layer of protoplanetary disks. The pressure gradient is the source of free energy that the instability taps into \citep{Youdin2007a} as it causes isolated gas to orbit with a sub-Keplerian speed, in contrast with the Keplerian rotation of dust in isolation\footnote{While the streaming instability as described by \citet{Youdin2005} springs from a gas pressure gradient, other causes for a discrepancy between the gas and dust orbital speeds like torques on the gas likewise lead to instability \citep{Lin2022}.}. The mutual drag between dust and gas rotating with different speeds results in a relative radial drift and instability. 

In addition to arguably being the leading candidate among mechanisms to induce planetesimal formation \citep[e.g.][]{Johansen2014, Lesur2023}, the streaming instability is by nature also a source of turbulence. The strength of both this turbulence and the dust diffusion it entails increases with the Stokes number, as do the length scales of turbulence (\citealt{Johansen2007a}, \citealt{Schreiber2018}, \citetalias{Schafer2020}, \citealt{Yang2021}, \citealt{Baronett2024}). This is in agreement with larger linear growth rates and wavelengths for higher Stokes number \citep{Youdin2005}. The instability drives largely isotropic turbulence when considering the radial and vertical dimensions (\citealt{Johansen2007a}, \citetalias{Schafer2020}, \citealt{Yang2021}, \citealt{Baronett2024}).

Both \citet{Schreiber2018} and \citet{Li2021} report that dust diffusion, on the other hand, is stronger radially than vertically, although this discrepancy disappears with increasing dust-to-gas ratio \citep{Schreiber2018, Gerbig2023}. In contrast, the work by \citet{Yang2021} and \citet{Baronett2024} reveals the opposite for Stokes numbers of order unity, and no anisotropy for smaller Stokes numbers. We note that of these studies only \citet{Li2021} includes the vertical stellar gravity. The dust scale height amounts to~\mbox{${\sim}1\%$} of the gas scale height (\citealt{Bai2010b}, \citealt{Yang2014}, \citealt{Li2018}, \citetalias{Schafer2020}) and is self-regulatory: If the dust would settle to a thinner layer, the density of this layer would be higher, and the instability would cause stronger turbulence and diffusion \citep{Bai2010b}.

Besides models of only the vertical shear instability or the streaming instability, we explore two scenarios in which both instabilities are active in \citetalias{Schafer2020}, \citetalias{Schafer2022}, and in this paper. In the scenario \emph{SIwhileVSI}, both instabilities begin to grow simultaneously. Here, turbulence in the mid-plane dust layer is predominantly caused by the streaming instability, and away from this layer by the vertical shear instability. In the scenario \emph{SIafterVSI}, on the other hand, the vertical shear instability saturates before the streaming instability starts to operate, and remains the primary source of turbulence both in and away from the dust layer. Nonetheless, the streaming instability causes turbulence locally in dust overdensities also in this scenario. Expanding on our previous study focused on gas turbulence, we aim to explore dust diffusion induced by the two instabilities in isolation and in conjunction in this paper.

This paper is structured as follows: We introduce our numerical model and its parameters in Sect.~\ref{sect:simulations}. In Section~\ref{sect:gas}, we examine gas turbulence induced by the vertical shear instability and the streaming instability individually and jointly, before commenting on the shape of the dust layer in Sect.~\ref{sect:dust_layer}. Sections~\ref{sect:vertical_diffusion} and~\ref{sect:radial_diffusion} are dedicated to a detailed investigation of vertical and radial dust diffusion, respectively. This is followed by a comparison between our models and the one presented by \citet{Flock2020}. We discuss implications and limitations of our work in Sect.~\ref{sect:discussion}, and conclude in Sect.~\ref{sect:conclusion}.

\section{Simulations}
\label{sect:simulations}
The numerical model employed in this study is the same as in \citetalias{Schafer2020} and \citetalias{Schafer2022}. We provide a summary of the model below, and refer to \citetalias{Schafer2020} for a more detailed description. 

\begin{table}
\caption{Simulation parameters}
\centering
\resizebox{\hsize}{!}{
\begin{tabular}{lcccccc}
\hline
\hline
Simulation name&Equation&Drag of dust&$t_{\textrm{d,init}}$ [$\kyr$]\tablefootmark{a}&$L_z$ [$H_{\gas}$]\tablefootmark{b}&$a$ [$\cm$]\tablefootmark{c}&$t_{\textrm{end}}$ [$\kyr$]\tablefootmark{d}\\
&of state&onto gas?&&&&\\
\hline
\hline
\textit{VSI\_a=0.3}&isothermal&no&$50$&$4$&$0.3$&$55$\\
\textit{VSI\_a=3}&isothermal&no&$50$&$4$&$3$&$55$\\
\hline
\textit{SI\_a=0.3}&adiabatic&yes&$0$&$2$&$0.3$&$5$\\
\textit{SI\_a=3}&adiabatic&yes&$0$&$2$&$3$&$2.5$\\
\hline
\textit{SIwhileVSI\_a=0.3}&isothermal&yes&$0$&$4$&$0.3$&$5$\\
\textit{SIwhileVSI\_a=3}&isothermal&yes&$0$&$4$&$3$&$10$\\
\hline
\textit{SIafterVSI\_a=0.3}&isothermal&yes&$50$&$4$&$0.3$&$55$\\
\textit{SIafterVSI\_a=3}&isothermal&yes&$50$&$4$&$3$&$60$\\
\hline
\hline
\end{tabular}
}
\tablefoot{
\tablefoottext{a}{Time after which particles representing the dust are initialised.}
\tablefoottext{b}{Vertical domain size, where~$H_{\rm g}$ is the gas scale height.}
\tablefoottext{c}{Dust particle size.}
\tablefoottext{d}{Time after which simulation ends.}
}
\label{table:simulations}
\end{table}

We applied the FLASH Code\footnote{\url{https://flash.rochester.edu/site/flashcode/}}\footnote{While we are not permitted to re-distribute the FLASH Code or any
of its parts, we are happy to share the modifications to the code that we
implemented to perform the simulations presented in this
paper upon request.} \citep{Fryxell2000} to simulate the gas and dust components of protoplanetary disks, the former on a Eulerian grid and the latter as Lagrangian particles. Our model includes the drag coupling between the two components as well as stellar gravity. In Table~\ref{table:simulations}, we list all simulations and the parameters that set them apart. The simulation names are composed of the scenario modelled and the dust size~$a$, the latter given in centimetres.

To begin with, we consider scenarios in which only either the vertical shear instability or the streaming instability is active. To ensure that the respective other instability does not operate, in the former scenario we consider only the drag exerted by the gas on the dust. (The drag of the dust onto the gas was included in all other scenarios.) In the latter scenario, the gas equation of state is adiabatic since under this condition the vertical shear instability is quenched by vertical buoyancy. (An isothermal equation of state was used in all other scenarios.) Moreover, we simulate the two instabilities in conjunction. In the scenario \emph{SIwhileVSI}, both the streaming instability and the vertical shear instability are in operation from the onset, while in scenario \emph{SIafterVSI} dust particles are introduced and the streaming instability therefore begins to grow only after~$50~\kyr$. At this point, the vertical shear instability has attained a saturated state everywhere in the simulation domain \citepalias{Schafer2020}.

\subsection{Domain size, resolution, and boundary conditions}
We performed two-dimensional, axisymmetric simulations with a cylindrical geometry, which is a natural choice to represent protoplanetary disks. The simulation domains extend from~$10~\au$ to~$100~\au$ in the radial dimension and to either~$1$ or~$2$ gas scale heights above and below the mid-plane in the vertical dimension. They are thus shaped like isosceles trapezoids with curved legs as the gas scale height increases non-linearly with the radius (see Eq.~\ref{eq:gas_scale_height}). We chose the larger vertical domain size for all models involving the vertical shear instability because we find this to be necessary for the instability to drive turbulence with a strength that is consistent with previous studies \citepalias{Schafer2020}. Since the streaming instability induces a dust scale height of~\mbox{${\sim}1\%$} of the gas scale height (\citealt{Bai2010b}, \citealt{Yang2014}, \citealt{Li2018}, \citetalias{Schafer2020}), the smaller vertical domain size is sufficient for our model of this instability only.

The base resolution of our simulations amounts to~$10$ grid cells per astronomical unit, or~$8$ and~$150$ cells per gas scale height in the mid-plane at the inner and outer radial domain boundaries, respectively (see Eq.~\ref{eq:gas_scale_height}). On top of that, we used static and adaptive mesh refinement with up to six levels of refinement. Every refinement level corresponds to a doubling in resolution, with the maximum resolution thus being equal to~$320$ cells per astronomical unit or~$270$ and~$4\,800$ cells per scale height at the radial boundaries. Static refinement was applied to increase the resolution by a factor of two within~$5~\au$ and by an additional factor of two (for a total factor of four) within~$1~\au$ above and below the mid-plane. Adaptive mesh refinement, on the other hand, enhanced the resolution of blocks consisting of~$10\times10$ cells if the number of particles in any cell within these blocks exceeded ten. Conversely, the resolution of a block was reduced if no particles remained in a cell. This approach to refinement enables us to simulate protoplanetary disks on a global scale, with a vertical domain extent that is sufficient to model the vertical shear instability, while simultaneously resolving diffusion in the dust mid-plane layer.

At the radial and vertical boundaries, gas and dust particles were allowed to move out of but not into the domains. To prevent the boundary conditions from artificially affecting our results, we excluded the innermost~$10~\au$ of the domains from all quantitative analysis. The vertical boundaries and the outer radial one are highly unlikely to influence our study of dust diffusion because the dust rapidly settles vertically and drifts radially away from these boundaries.

\subsection{Gas}
The gas is initially in vertical hydrostatic equilibrium, with the stellar gravity being balanced by a density gradient given by
\begin{equation}
\rho_{\gas}=\rho_{\gas}(z=0)\exp\left[-\frac{\gamma GM_{\textrm{S}}}{c_{\textrm{s}}^2}\left(\frac{1}{r}-\frac{1}{\sqrt{r^2+z^2}}\right)\right],
\end{equation}
where~$r$ and~$z$ are the radial and vertical coordinate, respectively, with~\mbox{$r=0$} and~\mbox{$z=0$} being the coordinates of the central star and the disk mid-plane, respectively. Furthermore,~$\gamma$ is the adiabatic index,~$G$ the gravitational constant,~\mbox{$M_{\textrm{S}}=1~M_{\sun}$} the stellar mass,~\mbox{$c_{\textrm{s}}=(\gamma RT/\mu)^{1/2}$} the sound speed,~$R$ the ideal gas constant,~$T$ the temperature, and~$\mu=2.33$ the mean molecular weight. The temperature initially does not vary with height. The gas scale height is equal to
\begin{equation}
H_{\gas}=\sqrt{\frac{c_{\textrm{s}}^2r^3(2\gamma GM_{\textrm{S}}-c_{\textrm{s}}^2r)}{(c_{\textrm{s}}^2r-\gamma GM_{\textrm{S}})^2}}=0.846~\au~\left(\frac{r}{10~\au}\right)^{5/4}.
\label{eq:gas_scale_height}
\end{equation}

The initial radial gas density and temperature gradients can be expressed as
\begin{equation}
\rho_{\gas}(z=0)=5.62\times10^{-12}~\g\,\cm^{-3}~\left(\frac{r}{10~\au}\right)^{-9/4}\text{ and}
\end{equation}
\begin{equation}
T=88.5~\K~\left(\frac{r}{10~\au}\right)^{-1/2}.
\end{equation}
The latter is adopted from the minimum mass solar nebula model \citep{Hayashi1981}, while the surface density gradient~\mbox{$\Sigma_{\gas}\propto r^{-1}$} is consistent with observed ones \citep{Andrews2009, Andrews2010}. The negative radial pressure gradient and the baroclinity resulting from the temperature gradient give rise to the streaming instability and the vertical shear instability, respectively \citep{Youdin2005, Nelson2013}.

As noted above, we employed an isothermal equation of state,
\begin{equation}
P=\frac{RT}{\mu}\rho_{\rm g},
\end{equation}
in models of the vertical shear instability in isolation or in coexistence with the streaming instability since under this condition there is no vertical buoyancy counteracting the vertical shear instability \citep{Nelson2013, Lin2015}. (In these models, the adiabatic index~\mbox{$\gamma=1$}.) Conversely, in our model of only the streaming instability buoyancy quenches the vertical shear instability since the equation of state is adiabatic,
\begin{equation}
P=K\rho_{\gas}^{\gamma},
\end{equation}
where~\mbox{$K=RT\rho_{\gas}^{1-\gamma}/\mu$} is the polytropic constant and the adiabatic index~\mbox{$\gamma=5/3$}. We note that the polytropic constant is constant in time only as it varies with density and temperature. The gas initially orbits with a sub-Keplerian velocity such that the centrifugal and the pressure gradient force balance the radial stellar gravity.

\subsection{Dust particles}
We adopted an approach to modelling dust aggregates in protoplanetary disks using Lagrangian particles that is commonly applied in simulations of the streaming instability \citep{Youdin2007a, Bai2010a}: The mass and momentum of each particle are equal to those of a huge number of aggregates, while the drag coupling to the gas is equal to that of a single aggregate. In every simulation,~$10^6$ particles were initialised with a radially uniform distribution, while their vertical positions were randomly sampled from a Gaussian distribution whose scale height amounts to~$10\%$ of the gas scale height. The noise in this vertical distribution constitutes the seed for the streaming instability. The particles were introduced at the start of simulations in our model of only the streaming instability and in the scenario~\emph{SIwhileVSI}, but not until after~$50~\kyr$ in our model of the vertical shear instability in isolation and in the scenario~\emph{SIafterVSI}. Our results are independent of whether~$5\times10^5$ or~$10^6$ particles were simulated and of the random seed, while the dependence on the initial scale height is examined in \citetalias{Schafer2020}.

All dust particles possess the same mass~\mbox{$m_{\dust}=2.53\times10^{24}~\g$}. Since the particles are initially uniformly distributed in the radial dimension, their mass can be calculated as
\begin{equation}
m_{\dust}=\frac{1}{N_{\dust}}\int_{L_r} 2\pi r\Sigma_{\dust}~{\rm d}r.
\end{equation}
Here,~\mbox{$N_{\dust}=10^6$} is the total particle number,~\mbox{$L_r=90~\au$} the radial domain size, and~$\Sigma_{\dust}$ the dust surface density. Because we set a constant ratio of dust to gas surface density and the latter is inversely proportional to the radius, the particle mass is radially constant. The surface ratio was chosen to amount to~$2\%$. We note that this value is twice as high as the canonical dust-to-gas ratio in the interstellar medium, but lies within the range of dust-to-gas mass ratios obtained from disk observations \citep[e.g.][]{Miotello2023}. The enhanced surface density ratio does not affect gas turbulence and dust diffusion in our model of the vertical shear instability in isolation since the drag of the dust onto the gas is not considered. Similarly, simulations of the streaming instability with a lower surface density ratio of~$1\%$ yield comparable turbulent strengths and dust scale heights to the~$2\%$ case \citepalias{Schafer2020}. Nonetheless, it is possible that dust clumping, which is stronger for higher surface density ratios, influences diffusion \citep{Li2021}. Furthermore, a higher dust-to-gas ratio results in a stronger effective buoyancy suppressing the vertical shear instability in the scenarios in which both instabilities coexist (\citealt{Lin2017, Lin2019}, \citetalias{Schafer2020}).

We conducted two simulations of every scenario, one with a dust size of~\mbox{$a=3~\mm$} and one with~\mbox{$a=3~\cm$}. These sizes are consistent with the maximum ones derived from the opacity spectral index of the thermal dust emission from protoplanetary disks \citep{Macias2019, Macias2021, CarrascoGonzalez2019, Tazzari2021a, Tazzari2021b, Mauco2021}. Because both dust sizes are less than the gas mean free path length everywhere in our model, that is to say Epstein drag is applicable, they can be converted to Stokes numbers in the disk mid-plane as
\begin{align}
\St(z=0)&=\frac{a\rho_{\rm s}}{c_{\rm s}\rho_{\rm g}(z=0)}\varOmega_{\rm K}(z=0)\nonumber\\
&=6\times10^{-3}~\frac{1}{\sqrt{\gamma}}\left(\frac{a}{3~\mm}\right)\left(\frac{r}{10~\au}\right).
\label{eq:Stokes_number}
\end{align}
Here,~\mbox{$\rho_{\rm s}=1~\g\,\cm^{-3}$} is the dust material density and~$\varOmega_{\textrm{K}}$ the Keplerian orbital frequency. This conversion relation is illustrated in Fig.~2 of \citetalias{Schafer2022}. The initial orbital velocity of the dust particles is Keplerian.

\section{Gas turbulence}
\label{sect:gas}
While the focus of this study is on dust diffusion, in this section we briefly discuss the gas turbulence that induces this diffusion. To this end, we provide a synopsis of the findings of \citetalias{Schafer2020} and complement these with new results.

In \citetalias{Schafer2020}, we report that the vertical shear instability causes anisotropic turbulence as the gas speeds are higher in the vertical than in the radial dimension \citep{Nelson2013, Stoll2017}. The Mach number of the former speeds amounts to~\mbox{$\mathcal{M}_{\gas,z}\approx0.1$} \citep{Flock2017}. The streaming instability, on the other hand, is a source of largely isotropic turbulence with a Mach number between~$0.001$ and~$0.01$ \citep{Johansen2007a, Flock2021, Li2021, Yang2021, Baronett2024}.

\begin{figure}[t]
  \centering
  \includegraphics[width=\columnwidth]{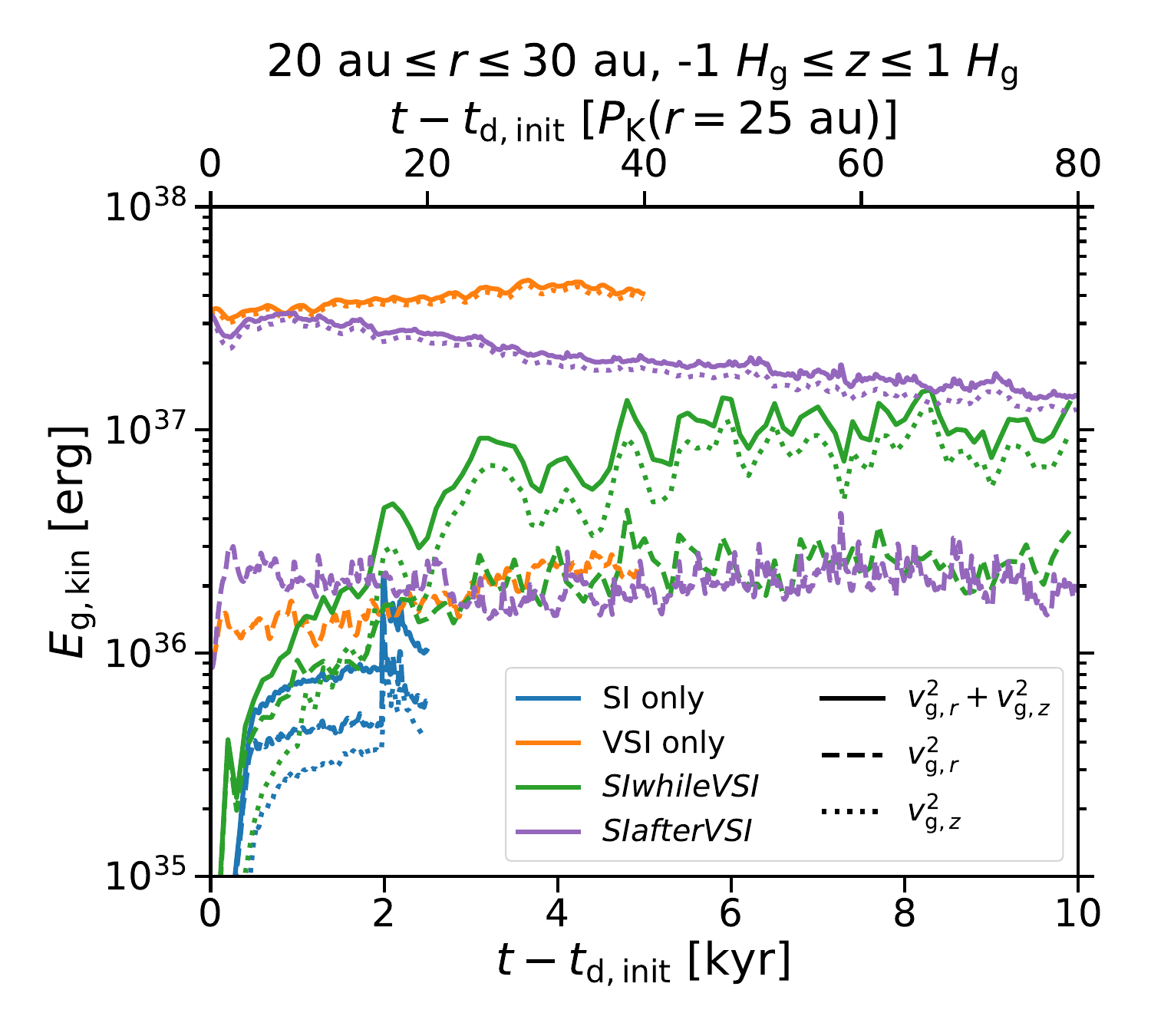} 
  \caption{Total gas kinetic energy~$E_{\textrm{g,kin}}$ in the region between~\mbox{$r=20~\au$} and~$30~\au$ and restricted to one gas scale height off the mid-plane as a function of time~$t$ after the initialisation of the dust particles at~$t_{\textrm{d,init}}$. Differently coloured lines represent simulations of the streaming instability and the vertical shear instability in isolation as well as of the scenarios \emph{SIwhileVSI} and \emph{SIafterVSI}, respectively, with the dust size being equal to~$3~\cm$ in all simulations. Like the model including only the vertical shear instability, the scenario \emph{SIafterVSI} as well as the scenario \emph{SIwhileVSI} at late times are characterised by a comparatively high total energy, with the energy of the vertical motions being substantially higher than that of the radial motions. In comparison, the total energy is less in the model of the streaming instability and in the scenario \emph{SIwhileVSI} at early times, but largely equally distributed among the radial and vertical velocity components.}
  \label{fig:kinetic_energy_comparison}
\end{figure}

These findings are reflected in Fig.~\ref{fig:kinetic_energy_comparison}. Here, we depict the evolution of the kinetic energy associated with the radial and vertical components of the gas velocity as well as the sum of the two components in all our scenarios. We note that the figure shows the evolution beginning with the time at which the dust particles are initialised, that is to say after the vertical shear instability has saturated in our model including only this instability and in the scenario \emph{SIafterVSI}.

 While the kinetic energy is overall greatest in our model of the vertical shear instability in isolation, it is more than an order of magnitude larger when considering only the vertical component than when taking into account only the radial one. In comparison, the streaming instability in isolation gives rise to turbulence with the least kinetic energy, with the energy of the radial gas motions being marginally larger than that of the vertical motions. It is interesting to note a spike in the kinetic energy after~\mbox{${\sim}2~\kyr$} in this model. We do not find a peak or a valley in the maximum or mean dust-to-gas density ratio at the same time, that is to say the abrupt increase in turbulent strength is not related to the instability suddenly inducing stronger or weaker dust concentration (see also \citetalias{Schafer2022}).

In the scenario \emph{SIwhileVSI}, both the streaming instability and the vertical shear instability are active from the outset. The gas kinetic energy is initially comparable to that in the model of the streaming instability in isolation, with the distribution among the radial and vertical velocity components being roughly equal. This shows that the streaming instability grows more rapidly in energy at least close to the mid-plane where the density is highest. Subsequently, though, the energy in the scenario \emph{SIwhileVSI} increases to exceed that in the model of the streaming instability only. Moreover, considerably more energy is associated with the vertical than the radial velocity component. That is, at later times the kinetic energy is dominated by the vertical shear instability. This is despite the turbulence in the dust mid-plane layer being primarily caused by the streaming instability \citepalias{Schafer2020}.

The vertical shear instability is the dominant source of gas turbulence in the scenario \emph{SIafterVSI}, both in and away from the mid-plane \citepalias{Schafer2020}. Thus, the total kinetic energy and the energies of the radial and vertical motions are similar to those in our model of the isolated vertical shear instability. Over time, though, particularly the total energy and the energy in the vertical velocities decline. We interpret this as a consequence of the settling dust causing an effective vertical buoyancy that weakens the vertical shear instability (\citealt{Lin2017}, \citealt{Lin2019}).

It is interesting to note that the kinetic energies in the scenarios \emph{SIwhileVSI} and \emph{SIafterVSI} attain comparable values~\mbox{${\sim}10~\kyr$} after dust initialisation. This is not true for the strength of dust diffusion especially in the vertical dimension, though, as can be seen from Fig.~5 of \citetalias{Schafer2022} and is detailed below. 

In Appendix~\ref{sect:power_spectra}, we present and analyse kinetic energy power spectra. We note, though, that their meaningfulness is limited by the grid in our simulations being neither periodic nor uniform.

\section{Dust layer thickness}
\label{sect:dust_layer}
\begin{figure*}[t]
  \centering
  \includegraphics[width=\textwidth]{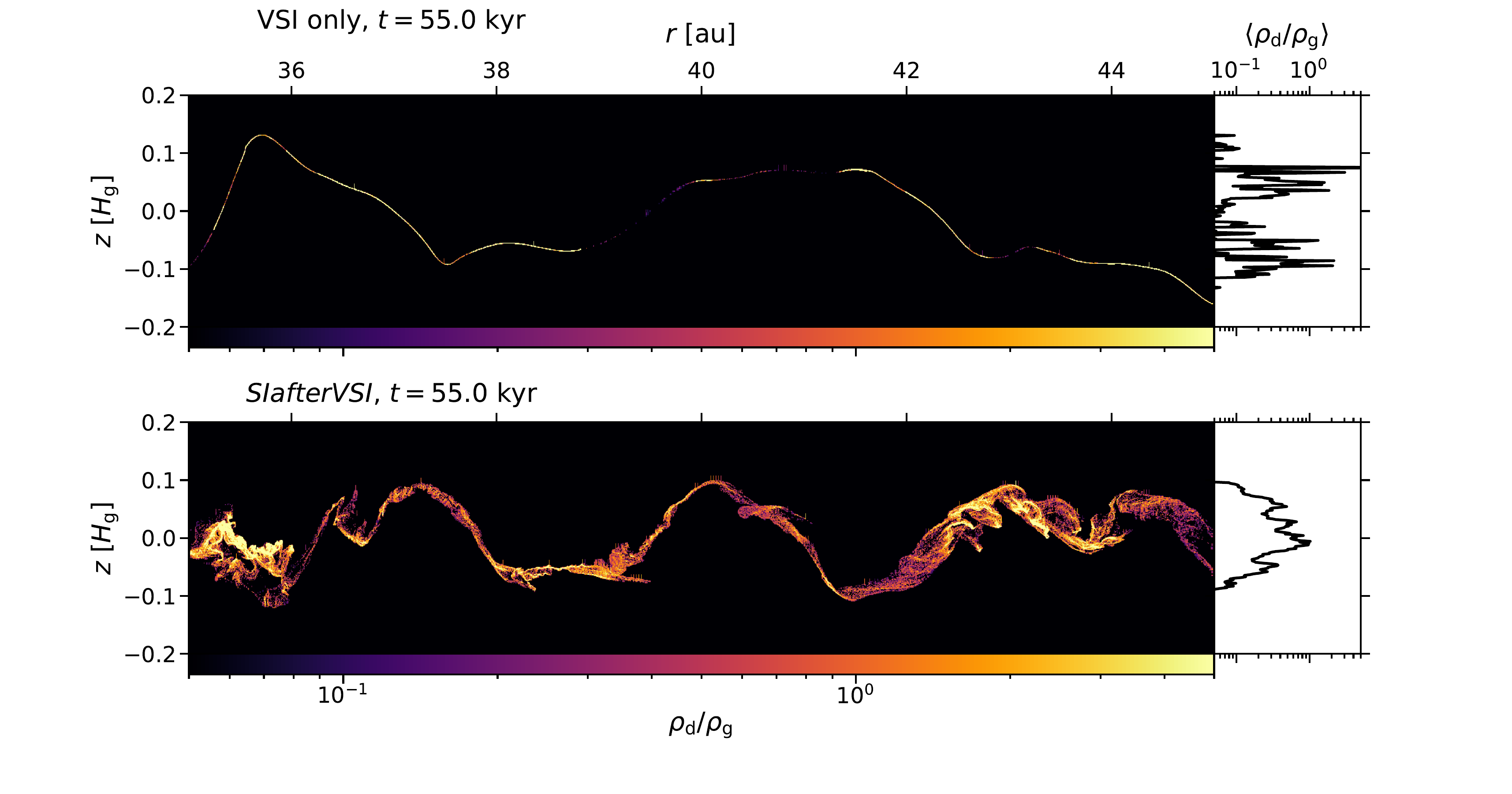} 
  \begin{minipage}{0.49\textwidth}
    \includegraphics[width=\textwidth]{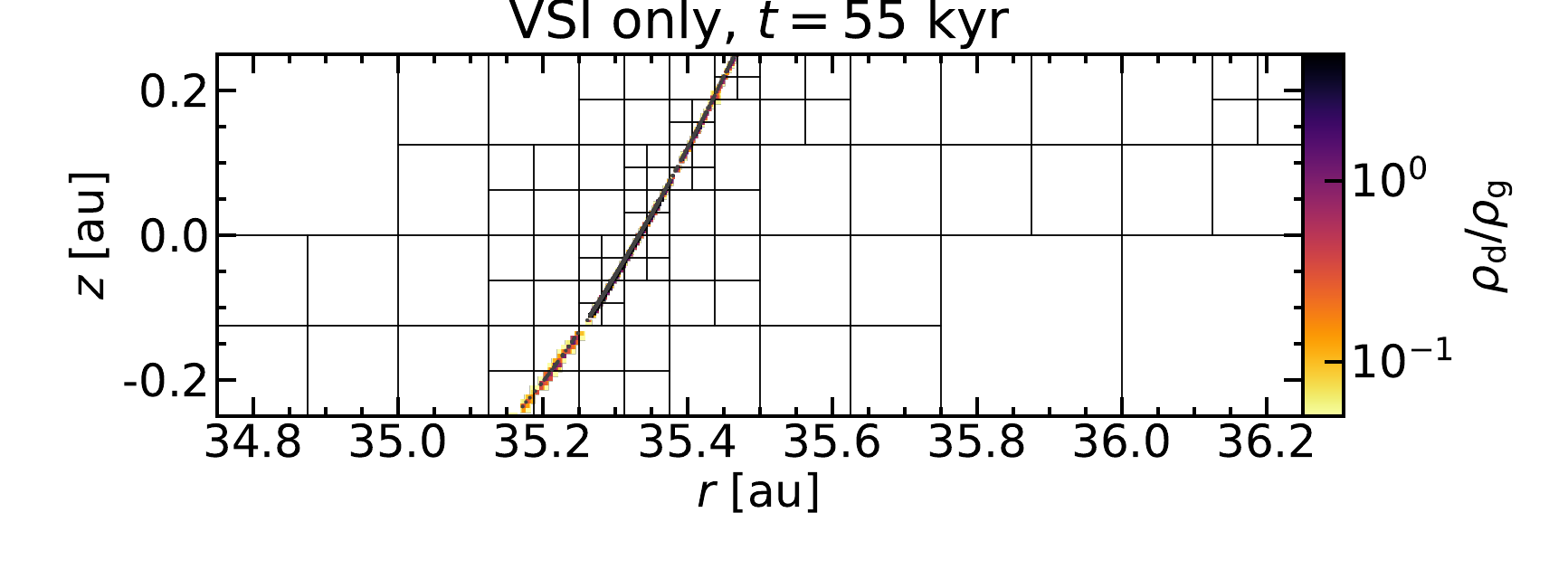}
  \end{minipage} 
  \hfill
  \begin{minipage}{0.49\textwidth}
    \centering
    \includegraphics[width=\textwidth]{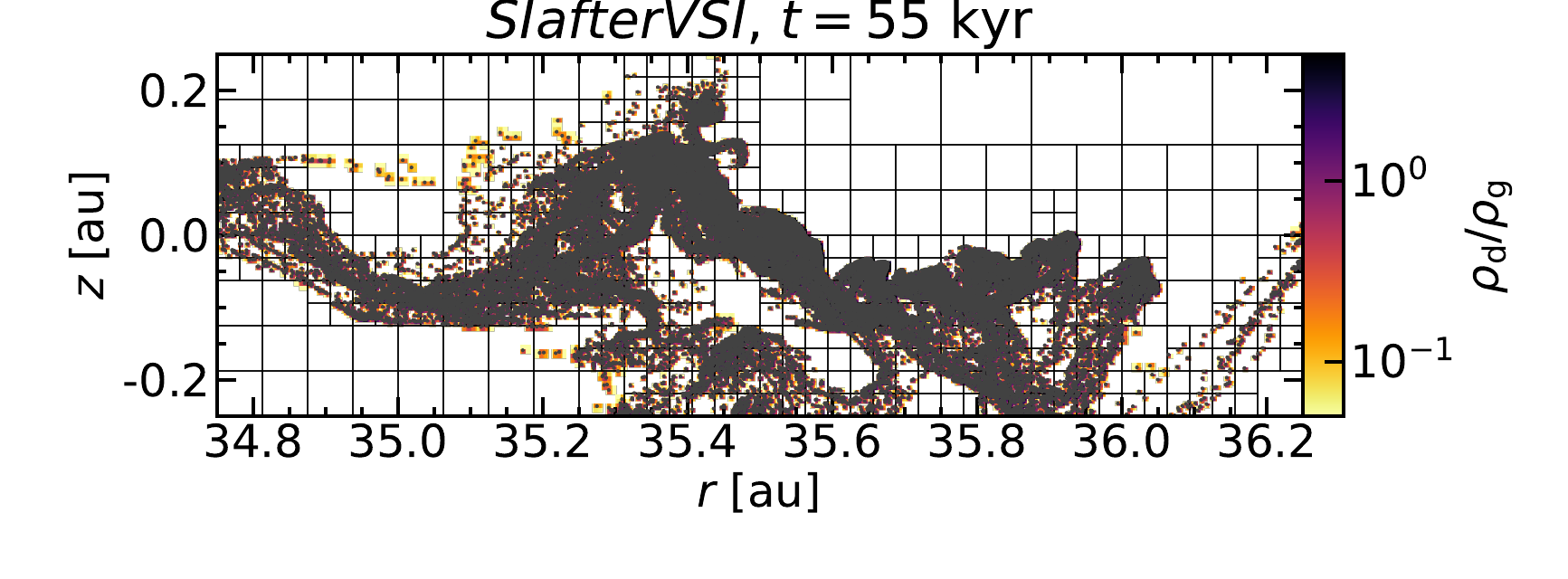} 
  \end{minipage} 
  \caption{Ratio of dust~$\rho_{\dust}$ to gas (volume) density~$\rho_{\gas}$ as a function of radius~$r$ and height~$z$, the latter in units of gas scale heights~$H_{\gas}$ (upper panels). Zoom-in on dust-to-gas density ratio~$\rho_{\dust}/\rho_{\gas}$ as a function of radius~$r$ and height~$z$, with grid structure and dust particles being plotted as black lines and grey dots, respectively (lower panels). Simulations of the vertical shear instability only and of the scenario \emph{SIafterVSI} with a dust size of~$3~\cm$ are depicted in the upper and left panel as well as lower and right panel, respectively. While the dust is stirred up to a marginally greater scale height  with respect to the mid-plane in the former simulation, the dust layer is much thinner. Its width indeed amounts to less than a grid cell, while the layer in the latter simulation extends over multiple cells both in the radial and in the vertical dimension.}
  \label{fig:dust_concentration_VSI_SIafterVSI}
\end{figure*}

In Figure~\ref{fig:dust_concentration_VSI_SIafterVSI}, we show the ratio of dust to gas density in our simulations of the vertical shear instability in isolation and of the scenario \emph{SIafterVSI} with a dust size of~$3~\cm$. The scale height of the dust, i.e.\ the height above or below the disk mid-plane to which the dust is elevated owing to large-scale turbulent diffusion, amounts to~\mbox{${\sim}10\%$} of the gas scale height in both simulations.

However, a stark contrast in the thickness of the dust layer is evident. The vertical shear instability alone causes a minuscule amount of small-scale diffusion both in the vertical and in the radial dimension, resulting in a dust layer that is not resolved in our model despite the application of adaptive mesh refinement. In comparison, in the scenario \emph{SIafterVSI} the dust layer is well-resolved because the streaming instability induces diffusion within the layer in this scenario.

It is important to note here that, although at first glance the much denser dust layer in our simulation of the vertical shear instability in isolation appears to provide better conditions for gravitational collapse and planetesimal formation, the drag exerted by the dust onto the gas is not included and the streaming instability thus not active in this model. However, neither this drag nor the streaming instability can be neglected when studying dust concentration and planetesimal formation.

In light of this finding, in the following we explore dust diffusion with a focus on its scale-dependence in all of our models. We begin by discussing the diffusion in the vertical dimension in the following section, with Sect.~\ref{sect:radial_diffusion} being dedicated to the radial diffusion. 

\section{Vertical diffusion}
\label{sect:vertical_diffusion}
\begin{figure*}[t]
  \begin{minipage}{0.49\textwidth}
    \centering
    \includegraphics[width=\textwidth]{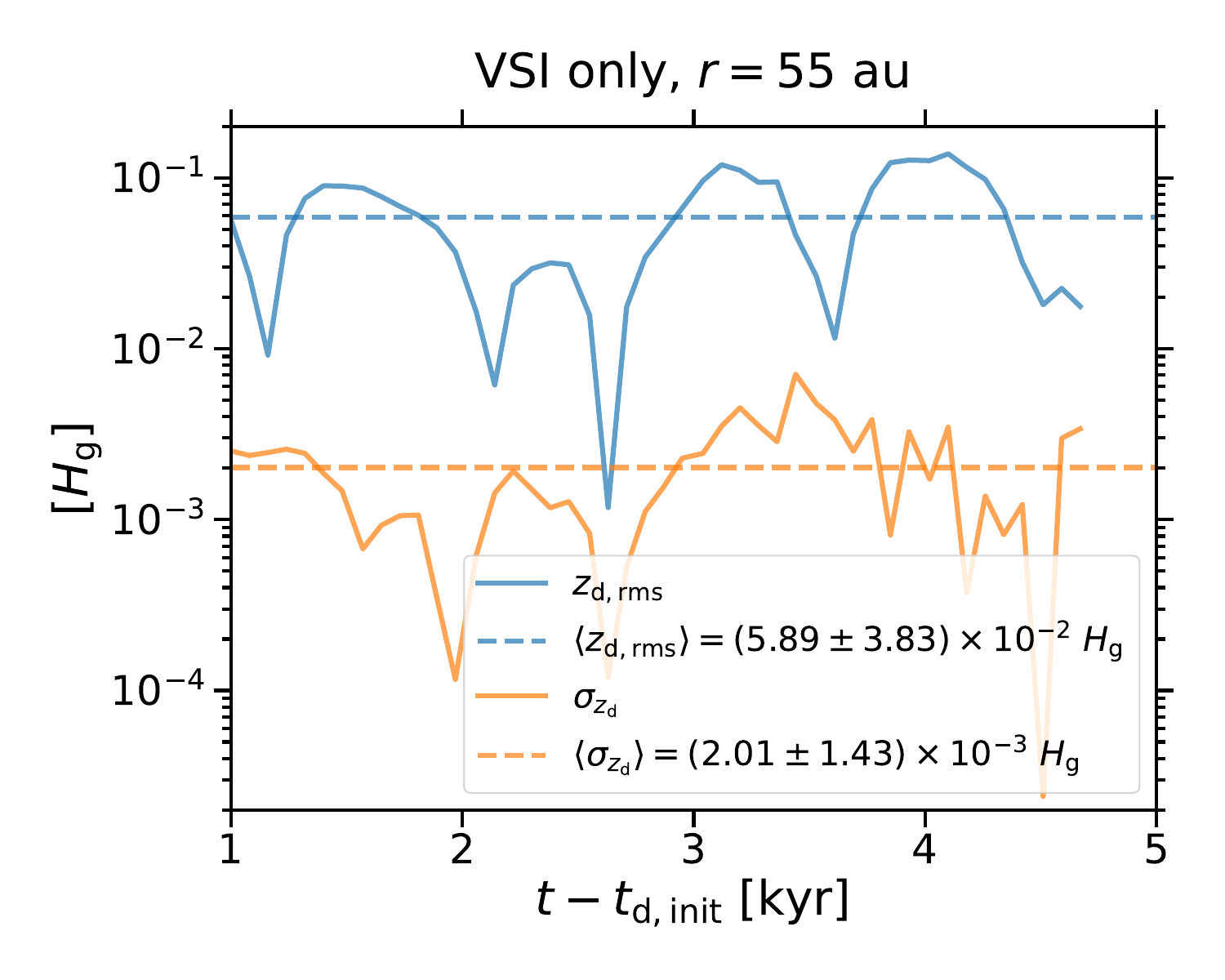}
  \end{minipage}
  \begin{minipage}{0.49\textwidth}
    \centering
    \includegraphics[width=\textwidth]{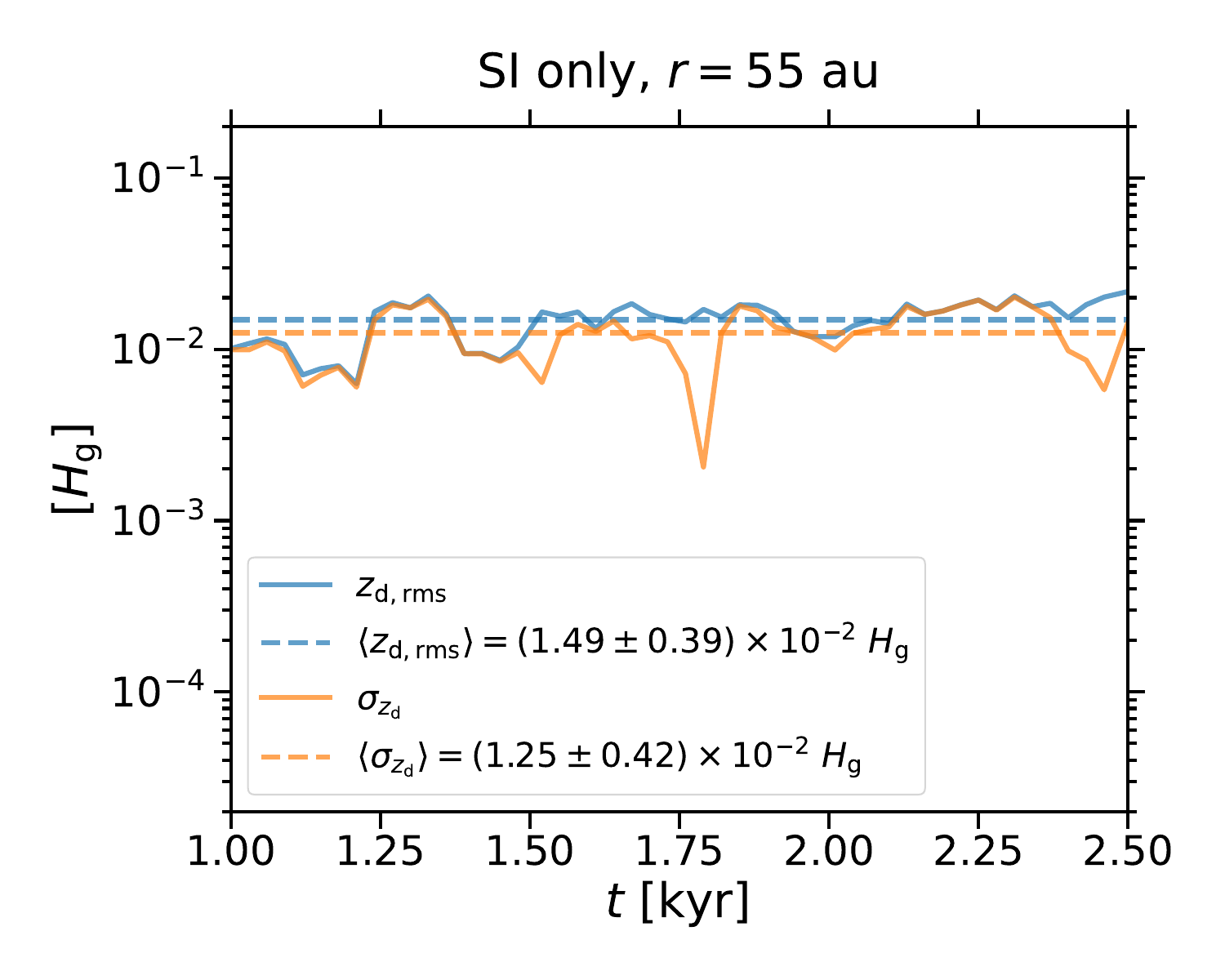}
  \end{minipage}
  \caption{Root mean square~$z_{\textrm{d,rms}}$ (blue solid line) and standard deviation~$\sigma_{z_{\dust}}$ (orange solid line) of the vertical dust particle positions as functions of time, considering only the time after dust settling. The mean of each quantity is shown as a dashed line. The left and right panel depict simulations of the vertical shear instability and the streaming instability with $3~\cm$-sized dust. The average root mean square is more than an order of magnitude greater than the standard deviation in former simulation, while they are similar in the latter simulation.}
  \label{fig:vertical_diffusion_VSI_SI}
\end{figure*}

To measure vertical diffusion on large and small scales, we employ the root mean square
\begin{equation}
z_{\textrm{d,rms}}=\sqrt{\langle z_{\dust}^2\rangle}
\end{equation}
and standard deviation
\begin{equation}
\sigma_{z_{\dust}}=\sqrt{\langle z_{\dust}^2\rangle-\langle z_{\dust}\rangle^2}
\end{equation}
of the vertical positions of the dust particles~$z_{\dust}$, i.e.\ the root mean square distances to the mid-plane and to the average vertical particle position, respectively. In other words, the former gives the dust scale height relative to the disk mid-plane, the latter with respect to the middle of the dust layer. Figure~\ref{fig:vertical_diffusion_VSI_SI} shows the evolution of these two quantities after the dust has sedimented for $3$-$\cm$-sized dust in our models of only the vertical shear instability and only the streaming instability.

As illustrated in the top and bottom left panels of Figure~\ref{fig:dust_concentration_VSI_SIafterVSI}, the vertical shear instability induces an almost threadlike, wave-shaped dust layer \citep[see also][]{Flock2017, Flock2020, Dullemond2022}. Accordingly, the time-averaged root mean square of the vertical dust positions is more than an order of magnitude larger than the mean standard deviation. In contrast, the streaming instability gives rise to a dust layer that possesses a Gaussian shape centred on the disk mid-plane \citep[see, e.g., the top panel of Fig.~3 of][]{Schafer2022, Bai2010b, Li2021}. Thus, the root mean square and standard deviation of the vertical dust positions are comparable.

\begin{figure*}[t]
  \centering
  \includegraphics[width=\textwidth]{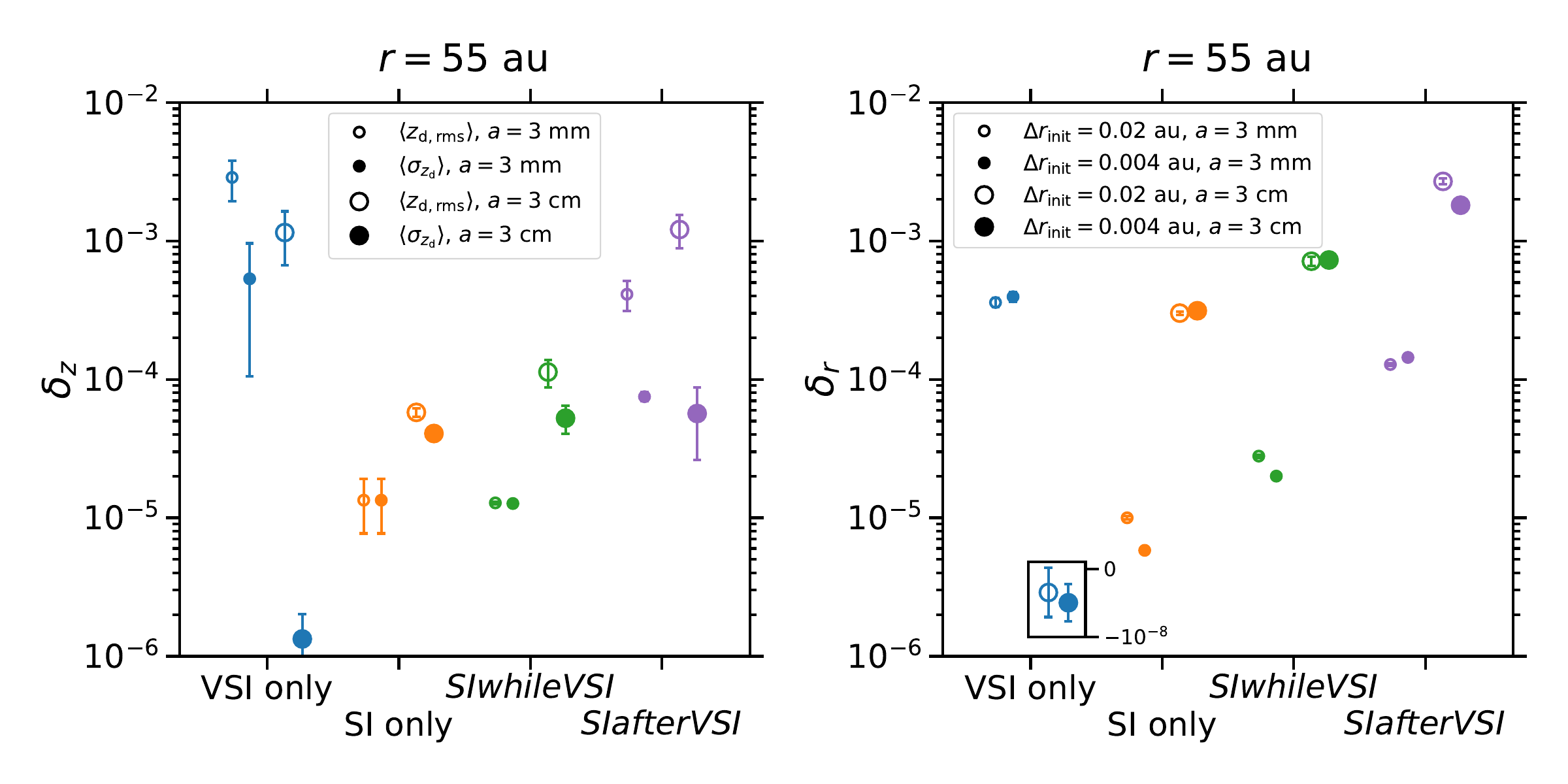} 
  \caption{Dimensionless coefficients of vertical dust diffusion~$\delta_z$ calculated from the time-averaged root mean square~$z_{\textrm{d,rms}}$ (circles) and standard deviation~$\sigma_{z_{\dust}}$ (squares) of the vertical dust particle positions (left panel). Dimensionless radial dust diffusion coefficients~$\delta_r$ computed considering all dust particles that are initially located within~$\Delta r_{\textrm{init}}=0.1~\au$ (circles) or~$\Delta r_{\textrm{init}}=0.01~\au$ (squares). In both panels, open and filled symbols represent simulations with a dust size of~$3~\mm$ and~$3~\cm$, respectively. Simulations of only the vertical shear instability, only the streaming instability, the scenario \emph{SIwhileVSI}, and the scenario \emph{SIafterVSI} are depicted using blue, orange, green, and purple markers, respectively. We note that the vertical diffusion coefficients in the vertical shear instability simulation with the larger dust size fall outside of the ordinate range and are therefore plotted in an inset with a different ordinate.}
\label{fig:dust_diffusion_comparison}
\end{figure*}

\begin{table*}
\caption{Dust diffusion parameters at~\mbox{$r=55~\au$}}
\centering
\resizebox{\hsize}{!}{
\begin{tabular}{lcccccc}
\hline
\hline
Scenario&$a$ [$\cm$]&$\delta_z(z_{\textrm{d,rms}})$\tablefootmark{a}&$\delta_z(\sigma_{z_{\dust}})$\tablefootmark{a}&$\delta_r(\Delta r_{\textrm{init}}=0.02~\au)$\tablefootmark{b}&$\delta_r(\Delta r_{\textrm{init}}=0.004~\au)$\tablefootmark{b}\\
\hline
\hline
VSI only&$0.3$&$(2.87\pm0.94)\times10^{-3}$&$(5.32\pm4.27)\times10^{-4}$&$(3.59\pm0.27)\times10^{-4}$&$(3.95\pm0.33)\times10^{-4}$\\
VSI only&$3$&$(1.15\pm0.48)\times10^{-3}$&$(1.33\pm0.67)\times10^{-6}$&$(-3.47\pm3.61)\times10^{-9}$&$(-4.98\pm2.72)\times10^{-9}$\\
\hline
SI only&$0.3$&$(1.34\pm0.57)\times10^{-5}$&$(1.34\pm0.57)\times10^{-5}$&$(1.00\pm0.03)\times10^{-5}$&$(5.82\pm0.25)\times10^{-6}$\\
SI only&$3$&$(5.77\pm0.40)\times10^{-5}$&$(4.06\pm0.46)\times10^{-5}$&$(3.01\pm0.08)\times10^{-4}$&$(3.13\pm0.12)\times10^{-4}$\\
\hline
\textit{SIwhileVSI}&$0.3$&$(1.28\pm0.02)\times10^{-5}$&$(1.27\pm0.03)\times10^{-5}$&$(2.79\pm0.07)\times10^{-5}$&$(2.00\pm0.11)\times10^{-5}$\\
\textit{SIwhileVSI}&$3$&$(1.13\pm0.26)\times10^{-4}$&$(5.24\pm1.19)\times10^{-5}$&$(7.13\pm0.53)\times10^{-4}$&$(7.28\pm0.52)\times10^{-4}$\\
\hline
\textit{SIafterVSI}&$0.3$&$(4.12\pm1.00)\times10^{-4}$&$(7.49\pm0.56)\times10^{-5}$&$(1.28\pm0.03)\times10^{-4}$&$(1.44\pm0.04)\times10^{-4}$\\
\textit{SIafterVSI}&$3$&$(1.21\pm0.33)\times10^{-3}$&$(5.66\pm3.04)\times10^{-5}$&$(2.69\pm0.13)\times10^{-3}$&$(1.81\pm0.12)\times10^{-3}$\\
\hline
\hline
VSI only\tablefootmark{c}&$0.1$&$(3.98\pm0.67)\times10^{-3}$&$(6.22\pm0.33)\times10^{-4}$&$(6.15\pm1.08)\times10^{-5}$&$(4.78\pm0.52)\times10^{-5}$\\
VSI only ($r=30~\au$)\tablefootmark{d}&$0.1$&$(8.96\pm0.38)\times10^{-4}$&$(7.41\pm0.19)\times10^{-4}$&$(7.72\pm0.79)\times10^{-4}$&$(4.98\pm1.25)\times10^{-4}$\\
\hline
\hline
\end{tabular}
}
\tablefoot{
\tablefoottext{a}{Dimensionless coefficient of vertical dust diffusion, calculated from either the time-averaged root mean square or the time-averaged standard deviation of the vertical particle positions at~\mbox{$r=55~\au$}.}
\tablefoottext{b}{Dimensionless radial dust diffusion coefficient, computed considering all dust particles that are initially located within~$0.02~\au$ or~$0.004~\au$ around~\mbox{$r=55~\au$}. Here,~$H_{\gas}$ and~$\varOmega_{\K}$ are the gas scale height and Keplerian orbital frequency.}
\tablefoottext{c}{Three-dimensional Pluto Code simulation presented by \citet{Flock2020}.}
}
\label{table:dust_diffusion}
\end{table*}

\citet{Youdin2007b} introduce an expression for the ratio of dust to gas scale height as a function of the dimensionless dust diffusion coefficient in the vertical dimension~$\delta_z$,
\begin{equation}
\frac{H_{\dust}}{H_{\gas}}=\sqrt{\frac{\delta_z}{\delta_z+\St}},
\end{equation}
where~$H_{\dust}$ is the dust scale height. Rearranging this equation, we can compute a diffusion coefficient both from the root mean square~$z_{\textrm{d,rms}}$,
\begin{equation}
\delta_z(z_{\textrm{d,rms}})=\frac{\St}{\left(\frac{H_{\gas}}{z_{\textrm{d,rms}}}\right)^2-1}
\end{equation}
and from the standard deviation~$\sigma_{z_{\dust}}$,
\begin{equation}
\delta_z(\sigma_{z_{\dust}})=\frac{\St}{\left(\frac{H_{\gas}}{\sigma_{z_{\dust}}}\right)^2-1},
\end{equation}
of the vertical particle positions. The resulting diffusion coefficients are depicted in the left panel of Figure~\ref{fig:dust_diffusion_comparison} and listed in the third and fourth columns of Table~\ref{table:dust_diffusion} for simulations of all four of our scenarios, one each with a dust size of~$3~\mm$ and of~$3~\cm$.

In line with what we discuss above and what can be seen in Fig.~\ref{fig:dust_concentration_VSI_SIafterVSI}, we find the vertical shear instability in isolation to induce considerably stronger large-scale than small-scale diffusion. This discrepancy is considerable in our simulation of centimetre-sized dust, where the diffusion coefficient derived from the root mean square of the dust positions is equal to~$10^{-3}$, but the one derived from the standard deviation to a minuscule~$10^{-6}$. The difference is less significant for the millimetre-sized dust that is more tightly coupled to the gas, sediments more slowly, and is thus more affected by the (large-scale) turbulence caused by the vertical shear instability. Here, both the large-scale diffusion coefficient amounts to~$3\times10^{-3}$ and the small-scale one to~$5\times10^{-4}$. 

The streaming instability alone, in contrast, in our model causes diffusion that is largely independent of scale. The diffusion coefficient computed from the root mean square vertical position of the centimetre-sized dust is marginally higher than the one obtained from the standard deviation, while for the millimetre-sized dust the two coefficients are in fact equal. All four diffusion coefficients are of the order of~$10^{-5}$, with the two coefficients of the larger dust being greater by a factor of a few than those of the smaller dust. This is in agreement with both linear analysis and previous non-linear simulations showing that the streaming instability causes stronger turbulence and diffusion if the Stokes number of the dust is higher (\citealt{Youdin2005}, \citealt{Johansen2007a}, \citealt{Schreiber2018}, \citetalias{Schafer2020}, \citealt{Yang2021}, \citealt{Baronett2024}).
Nonetheless, \citealt{Schreiber2018} report a positive correlation also between diffusion coefficient and scale. We caution, though, that their simulations do not include the vertical stellar gravity. 

Since the streaming instability is the dominant source of turbulence and diffusion in the dust layer of the scenario \emph{SIwhileVSI} (\citetalias{Schafer2020}), the diffusion coefficients in this scenario are comparable to the ones arising if only the streaming instability is considered. Nevertheless, for the centimetre-sized dust the diffusion coefficient calculated from the standard deviation of the vertical positions, and even more so the one derived from the root mean square, are greater in the \emph{SIwhileVSI} simulation than in the simulation of the streaming instability only. We interpret this as the streaming instability sufficiently diffusing the dust for it to be susceptible to the large-scale diffusion induced by the vertical shear instability.

Similar to in our model of the vertical shear instability alone, in the scenario \emph{SIafterVSI} small-scale diffusion is significantly weaker than large-scale diffusion. This is because in this scenario turbulence and diffusion are mainly driven by the vertical shear instability also in the dust layer \citepalias{Schafer2020}. Nonetheless, since the scenario \emph{SIafterVSI} includes the drag exerted by the dust on the gas, the vertical shear instability is suppressed by an effective buoyancy resulting from the tendency of the dust to settle to the disk mid-plane (\citealt{Lin2019}, \citetalias{Schafer2020}). The diffusion coefficients of the millimetre-sized dust are thus reduced compared to when only the vertical shear instability is simulated. On the other hand, we find the streaming instability to contribute to the turbulence in the dust layer (\citetalias{Schafer2020}), and indeed to compensate for the lack of small-scale diffusion caused by the vertical shear instability. This is evident both from the middle and bottom right panels of Fig.~\ref{fig:dust_concentration_VSI_SIafterVSI} and from the coefficient of small-scale diffusion of the centimetre-sized dust, that is to say the coefficient obtained from the standard deviation of the vertical dust positions, being similar to the value in the scenario \emph{SIwhileVSI}.

\section{Radial diffusion}
\label{sect:radial_diffusion}
\begin{figure*}[t]
  \begin{minipage}{0.49\textwidth}
    \centering
    \includegraphics[width=\textwidth]{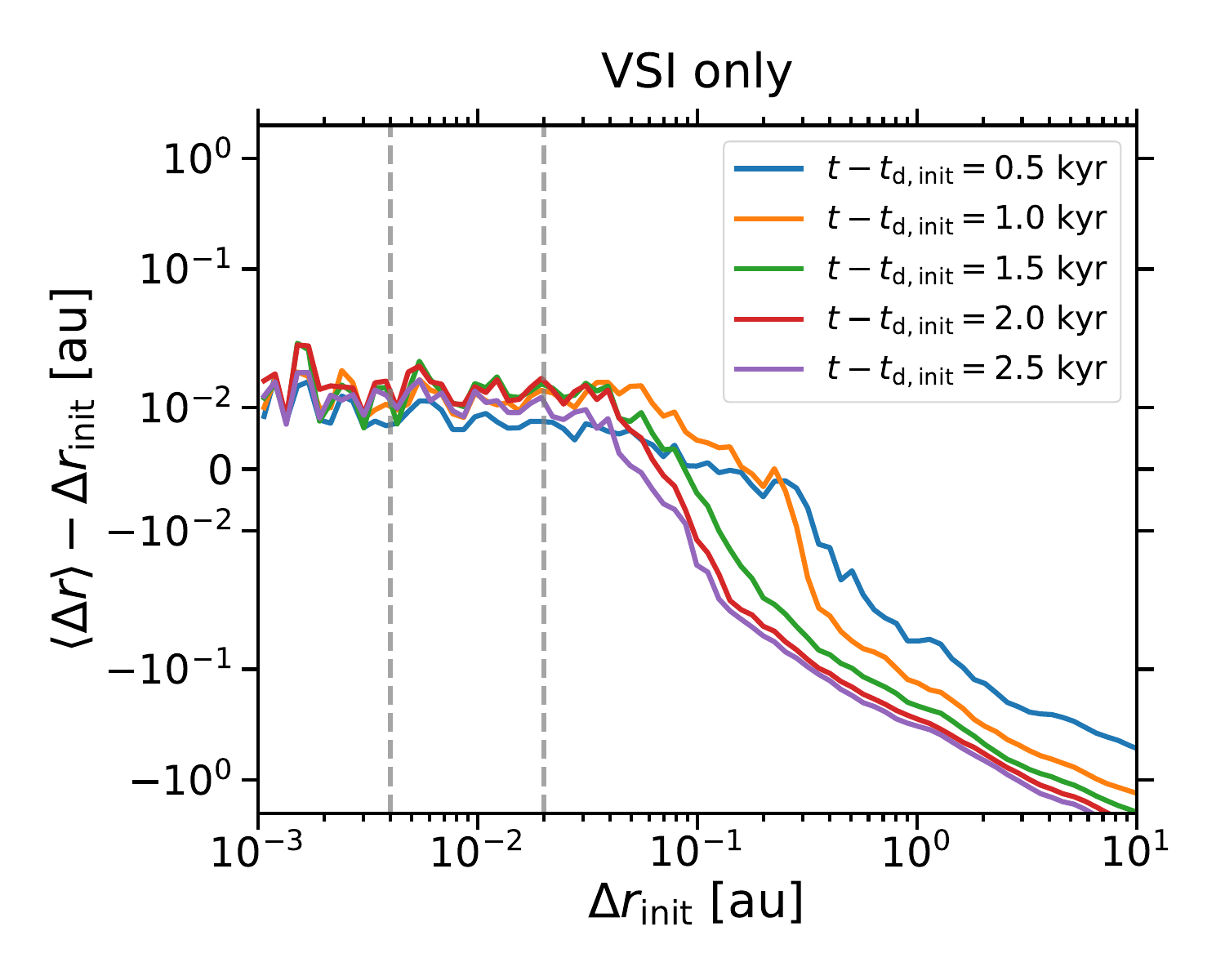}  
  \end{minipage} 
  \hfill
  \begin{minipage}{0.49\textwidth}
    \centering
    \includegraphics[width=\textwidth]{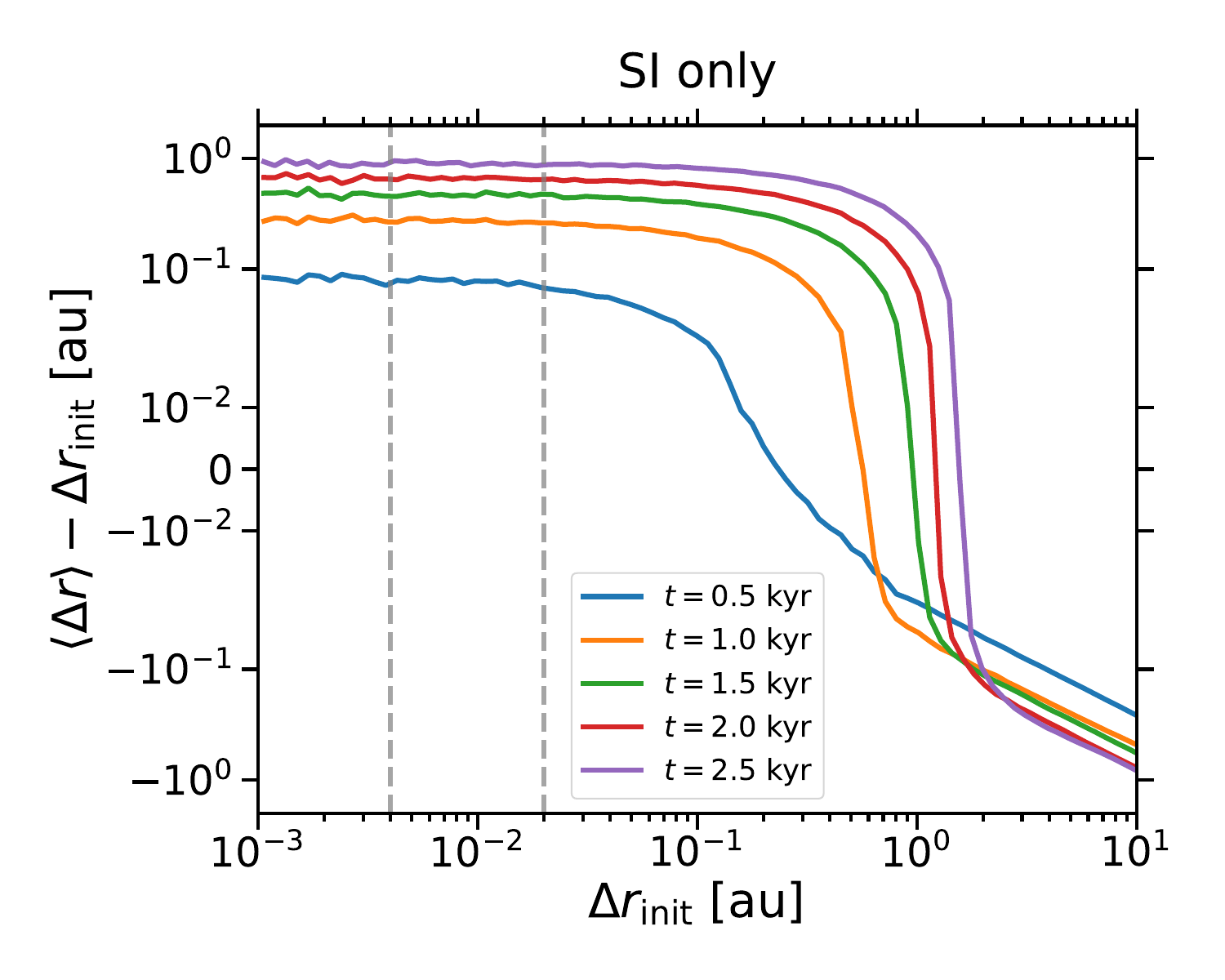} 
  \end{minipage} 
  \caption{Difference between average and initial radial separation~$\Delta r$ of dust particle pairs as a function of their initial separation in simulations of the vertical shear instability (left panel) and the streaming instability (right panel) with $3~\cm$-sized dust.  We note that the scale of the abscissa is linear between~$10^{-2}$ and~$-10^{-2}$ and logarithmic otherwise. If their initial separation is less than~\mbox{${\sim}0.1~\au$} (equivalent to~$0.08$ and $1.5$ gas scale heights at the inner and outer radial boundaries), the distance between two particles remains roughly constant if only the vertical shear instability is simulated, while the streaming instability causes the distance to gradually increase with time (differently coloured solid lines). The dashed grey lines mark initial separations of~$0.004~\au$ and~$0.2~\au$; we choose these scales to measure small-scale and large-scale radial diffusion induced by the two instabilities individually and jointly. If the initial separation exceeds~\mbox{${\sim}0.1~\au$}, on the other hand, two particles move closer together with time in both simulations. This is because the inwards drift speed of the particles increases with the radius, and the separation of particles at different radii thus decreases.}
\label{fig:radial_distance_diffusion_VSI_SI}
\end{figure*}

To examine radial dust diffusion induced by the streaming instability and the vertical shear instability in isolation and in conjunction, we first need to distinguish the scales on which turbulence is the main source of diffusion from the scales on which it is primarily caused by the inwards drift that arises even in the absence of turbulence. To this end, we track the separation of pairs of dust particles through time, specifically of all pairs among~$10,000$ particles randomly selected in every simulation. Figure~\ref{fig:radial_distance_diffusion_VSI_SI} shows the evolution of the separation of pairs of centimetre-sized dust particles in simulations of the vertical shear instability and the streaming instability, respectively. It can be gathered that, if their initial separation is large, two dust particles approach each other over time. This is because the outer particle drifts more rapidly inwards than the inner particle since the radial drift speed scales with the dust Stokes number, which in turn is proportional to the dust size (see Eq.~\ref{eq:Stokes_number} and Sect.~4.2 of \citetalias{Schafer2022}).

If it initially is less than~\mbox{${\sim}0.1~\au$} -- this corresponds to~$0.08$ gas scale heights at the inner and~$1.5$ scale heights at the outer radial domain boundary -- on the other hand, in the simulation of only the vertical shear instability the pair separation remains largely constant. Consistent with what can be seen from Fig.~\ref{fig:dust_concentration_VSI_SIafterVSI} and what we discuss in the previous section with respect to the vertical dimension, this demonstrates that the vertical shear instability induces minimal diffusion in the radial dimension. In contrast, the diffusion induced by the streaming instability in isolation leads to the pair separations gradually increasing with time, with this increase being largely independent of the initial separation.

To investigate the scale-dependence of radial diffusion, in what follows we compute and compare diffusion coefficients on scales of~$0.02~\au$ and of~$0.004~\au$. As is evident from Fig.~\ref{fig:radial_distance_diffusion_VSI_SI}, the former is the largest scale on which the effect of the inwards drift on the relative diffusion of the pairs of centimetre-sized dust particles in our simulation of the vertical shear instability is negligible -- since turbulence-induced diffusion is stronger, the transition to drift-induced diffusion occurs on greater scales in all other simulations. Our choice of the smaller scale, on the other hand, is constrained by the simulation resolution, as the grid cell edge length at the highest level of refinement amounts to~$3.125\times10^{-3}~\au$. 

The dust diffusion coefficient in the radial dimension can be calculated as \citep{Youdin2007b, Johansen2007a}
\begin{equation}
D_r=\frac{1}{2}\frac{\Delta \sigma_r^2}{\Delta t},
\end{equation}
where~$\Delta\sigma_r^2/\Delta t$ denotes the temporal change of the standard deviation of the radial dust particles positions\footnote{While the evolution of the mean particle position reflects their inwards drift, the standard deviation is determined by the diffusion owing to turbulence.}, and be non-dimensionalised as~\mbox{$\delta_r=D_r/(H_{\gas}^2\varOmega_{\textrm{K}})$}. To obtain coefficients for the larger and the smaller scale established in the previous paragraph, all particles that are initially located within~$0.02~\au$ and~$0.004~\au$, respectively, around the radial midpoint of our simulation domains at~$55~\au$ are taken into account. We perform a linear regression of the squared standard deviation of their radial coordinates as a function of time, only taking into consideration the time after the dust sedimentation has concluded. The two dimensionless diffusion coefficients derived in this manner for every simulation are shown in the right panel of Fig.~\ref{fig:dust_diffusion_comparison} and listed in the fifth and sixth columns of Table~\ref{table:dust_diffusion}.

As is indicated by Figure~\ref{fig:radial_distance_diffusion_VSI_SI}, both the streaming instability and the vertical shear instability, separately or jointly, induce radial dust diffusion that depends not or only marginally on scale. This is in agreement with what \citet{Schreiber2018} observe in their two-dimensional simulations of the streaming instability. Nevertheless, we note that the radial scales we consider differ only by a factor of five, and a dependence may emerge in models resolving a broader range of scales.

As noted above, in our model the centimetre-sized dust experiences negligible diffusion by the vertical shear instability -- the dimensionless diffusion coefficients in this case are of the order of~$10^{-9}$ and in fact negative. The coefficients in the simulation including millimetre-sized dust, on the other hand, amount to~\mbox{$4\times10^{-4}$}, similar to the coefficient of the small-scale vertical diffusion. A consistent picture of dust diffusion caused by the vertical shear instability emerges, reflected in Fig.~\ref{fig:dust_concentration_VSI_SIafterVSI}: The diffusion is strongest on large scales in the vertical dimension, causing an undulating dust layer. Within this layer, though, the diffusion is isotropic and considerably weaker, by an order of magnitude for millimetre-sized dust while negligible if the dust is centimetre-sized.

In our model of the streaming instability, on the other hand, whether or not diffusion is anisotropic is dependent on the dust size. The simulation of millimetre-sized dust yields diffusion coefficients of~$10^{-5}$ both in the radial and the vertical dimension. In contrast, we obtain vertical coefficients of~${\sim}5\times10^{-5}$ but radial coefficients of~$3\times10^{-4}$, larger by a factor of~$6$, from the simulation of centimetre-sized dust. This is consistent with the findings by \citet{Schreiber2018}, who show that dust is in general more strongly diffused in the radial than in the vertical dimension if its Stokes number amounts to~$0.1$ (see their Figs.~10 and~26), but the diffusion is mostly isotropic if the Stokes number is equal to~$0.01$ (see their Fig.~20). (These Stokes numbers approximately correspond to the dust sizes in our simulations, as can be gathered from Eq.~\ref{eq:Stokes_number}.) A similar trend in the relation between anisotropy and Stokes number emerges in the study by \citet{Li2021}, although the ratio between radial and vertical diffusion coefficients amounts to~$5$ in both their simulations of dust with a Stokes number of~$0.01$ and with a Stokes number of~$0.1$ (see their Fig.~6). We note that both \citet{Schreiber2018} and \citet{Gerbig2023} find this anisotropy to vanish in dust overdensities where the dust-to-gas ratio is high. Additionally, \citet{Yang2021} and \citet{Baronett2024} find diffusion to be isotropic for the Stokes numbers we study, and to be stronger in the vertical than in the radial dimension if the Stokes numbers is of order unity.

All radial diffusion coefficients, both of the millimetre-sized and the centimetre-sized dust as well as both on the smaller and on the larger scale, are doubled to tripled in the scenario \emph{SIwhileVSI} compared to our model of the streaming instability in isolation. This indicates that, even though the streaming instability is the main source of turbulence in the dust layer causes turbulence locally in dust overdensities \citepalias{Schafer2020}, the vertical shear instability contributes to the dust diffusion. Furthermore, the contribution is independent of dust size and scale. This is noticeable since this instability induces virtually no diffusion of the centimetre-sized dust when regarded alone. Similar to what we discuss with regards to vertical diffusion in this scenario, we interpret this as the streaming instability giving rise to smaller-scale diffusion that is picked up on by larger-scale diffusion owing to the vertical shear instability.

In comparison with the scenario \emph{SIwhileVSI}, in the scenario \emph{SIafterVSI} the radial diffusion coefficients are again enhanced by a factor of~$3-7$. They amount to~$10^{-4}$ for the millimetre-sized dust, less than when the vertical shear instability alone is considered but higher than when only the streaming instability is simulated. In the scenario \emph{SIafterVSI}, the vertical shear instability is the dominant source of turbulence in the dust layer, but the streaming instability causes turbulence locally in dust overdensities \citepalias{Schafer2020}. It is thus plausible that by taking into account all dust particles in a certain radius range, we infer coefficients that represent an ``average'' over the dust diffusion induced by the streaming instability inside overdensities and by the vertical shear instability outside of them.

Nonetheless, in our simulation of the scenario \emph{SIafterVSI} with centimetre-sized dust, the radial diffusion coefficients are as large as~${\sim}2\times10^{-3}$, slightly higher even than the large-scale diffusion coefficient in the vertical dimension. This implies that the streaming instability and the vertical shear instability jointly diffuse dust of this size in the radial dimension. As in the scenario \emph{SIwhileVSI}, this can be explained by the streaming instability inducing a base level of diffusion that renders the dust susceptible to the large-scale diffusion by the vertical shear instability. 

\section{Comparison with three-dimensional model of vertical shear instability}
\label{sect:comparison}
Since the inability of the vertical shear instability to diffuse centimetre-sized dust in the radial dimension as well as on small scales in the vertical dimension is unexpected in our eyes, we aim to validate of our model of this instability. To this end, in this section we compare it with the model of the instability presented in \citet[][hereafter F20]{Flock2020}. We note that the dust included in the latter authors' model is at most~$1~\mm$ in size, and a direct comparison is thus possible only with our simulation of dust with a size of~$3~\mm$. -- Since \citetalias{Flock2020} consider a dust material density of~$3~\g\,\cm^{-3}$ while we assume a material density of~$1~\g\,\cm^{-3}$, the Stokes number of the dust is indeed identical in the two models if the properties of the gas are. Nonetheless, a verification of our simulation of millimetre-sized dust can be extended also to the simulation of centimetre-sized dust.

\begin{figure*}[t]
  \centering
  \includegraphics[width=\textwidth]{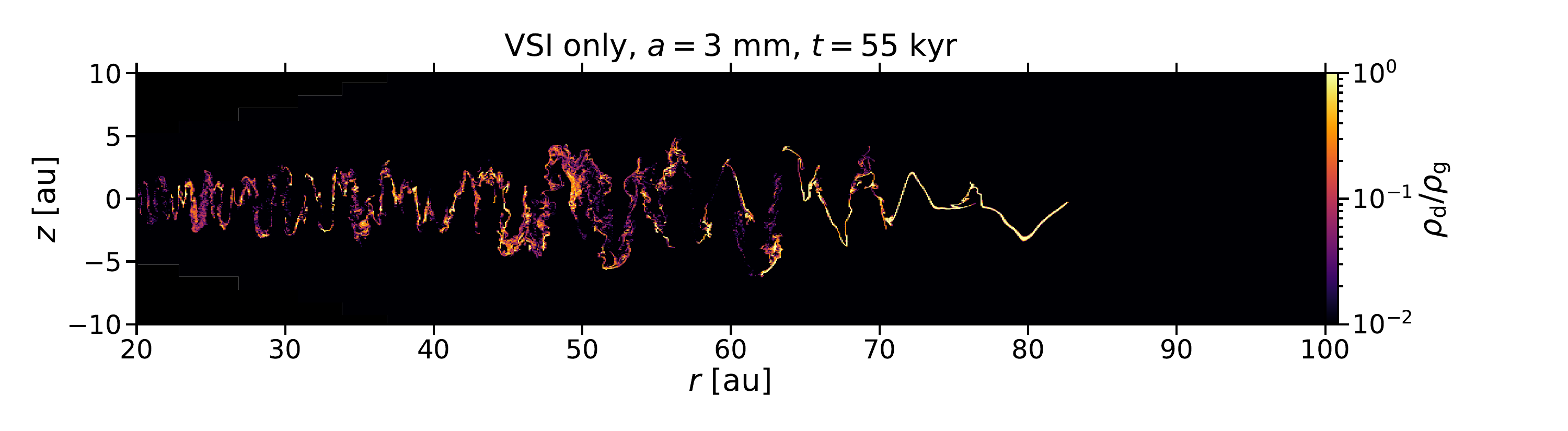} 
  \caption{Dust-to-gas density ratio~$\rho_{\dust}/\rho_{\gas}$ as a function of radius~$r$ and height~$z$ in our two-dimensional simulation of the vertical shear instability with~$3~\mm$-sized dust. The dust layer is similar in thickness, both with respect to mid-plane of the disk and to the mid-plane of the layer itself, to the layer of millimetre-sized dust in the model by \citetalias{Flock2020} (see their Fig.~D1).}
  \label{fig:dust_concentration_VSI}
\end{figure*}

\citetalias{Flock2020} investigate a three-dimensional radiation hydrodynamical simulation of an irradiated T Tauri star-disk system extending from~$20$ to $100~\au$. The grid resolution is uniform with around~$70$ cells per gas scale height. Half a million dust particles each of $1~\mm$ and of~$100~\mum$ were embedded to study their motions. Since the drag exerted by the particles on the gas was neglected, the streaming instability is not active. Similar to us, these authors observe mixing of the dust owing to the vertical shear instability as well as radial drift. While at first the mm-sized dust is concentrated in a narrow layer similar to what is shown in Fig.~\ref{fig:dust_concentration_VSI}, as the instability grows the dust layer evolves to be broader in both the radial and the vertical dimension.

Figure~\ref{fig:dust_concentration_VSI} depicts the dust layer in our simulation of solely the vertical shear instability and mm-sized dust. As is evident when comparing to Fig.~D1 of \citetalias{Flock2020}, both the scale height relative to the disk mid-plane and the thickness of this layer is comparable in our and in their simulation.

\begin{figure*}[t]
  \centering
  \includegraphics[width=\textwidth]{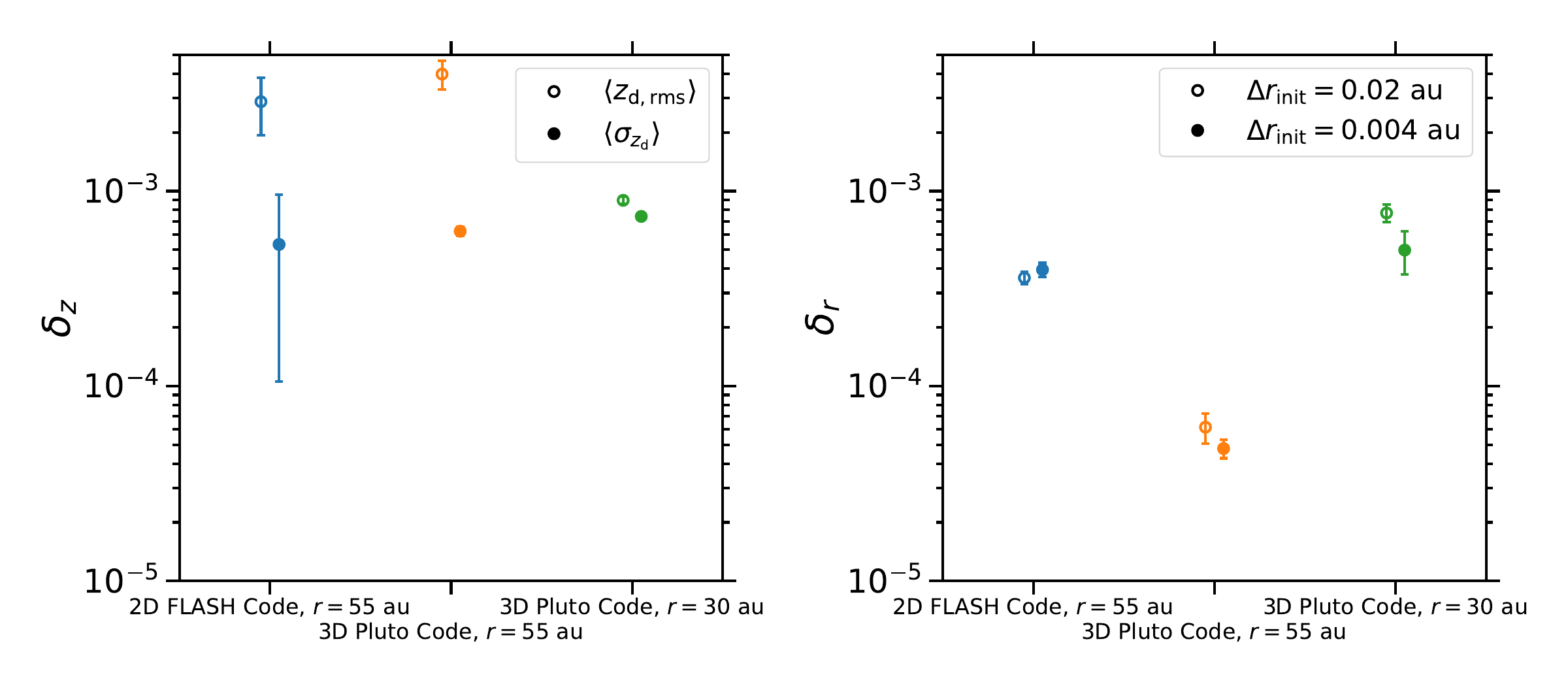} 
  \caption{\emph{Left panel:} Average root mean square~$z_{\textrm{d,rms}}$ and standard deviation~$\sigma_{z_{\dust}}$ of vertical particle positions. \emph{Right panel:} Radial dust diffusion coefficients~$D_r$ for~\mbox{$\Delta r_{\textrm{init}}=0.1~\au$} (circles) and~\mbox{$\Delta r_{\textrm{init}}=0.01~\au$} (squares). In both panels, blue, orange, and green symbols represent quantities at~\mbox{$r=55~\au$} in our two-dimensional simulation of the vertical shear instability with a dust size of~$3~\mm$ as well as at~\mbox{$r=55~\au$} and at~\mbox{$r=30~\au$} for the dust with a size of~$1~\mm$ in the three-dimensional simulation conducted by \citetalias{Flock2020}. While the vertical diffusion coefficients at~\mbox{$r=55~\au$} in the latter simulation and in ours are comparable, the radial coefficients are larger by almost an order of magnitude in our simulation. Nonetheless, vertical diffusion is similarly dependent on scale while radial diffusion is scale-independent both in the two- and the three-dimensional simulation. At~\mbox{$r=30~\au$}, the Rossby wave instability induces a giant vortex in the simulation by \citetalias{Flock2020}, which gives rise to diffusion that is similar in strength in the radial and vertical dimensions as well as on small and large scales.}
\label{fig:dust_diffusion_comparison_VSI}
\end{figure*}

This observation by eye is corroborated by the dust diffusion coefficients we infer. Figure~\ref{fig:dust_diffusion_comparison_VSI} depicts and Table~\ref{table:dust_diffusion} lists the dimensionless coefficients measured at radii of~$30~\au$ and~$55~\au$ in the simulation by \citetalias{Flock2020} as well as at~\mbox{$r=55~\au$} in our simulation. These coefficients are computed in the same manner as the ones shown in Fig.~\ref{fig:dust_diffusion_comparison}, that is they represent large-scale and small-scale diffusion in the vertical and the radial dimension. Indeed, the two vertical diffusion coefficients (see the left panel of Fig.~\ref{fig:dust_diffusion_comparison_VSI}) calculated at~\mbox{$r=55~\au$} are similar in our two-dimensional and their three-dimensional simulation. This concordance indicates that the results obtained from our model are robust, particularly with respect to the dust scale height, despite it being axisymmetric.

The coefficients of radial diffusion (right panel of Fig.~\ref{fig:dust_diffusion_comparison_VSI}), on the other hand, are smaller by almost an order of magnitude in the simulation by \citetalias{Flock2020} than in our simulation. This potentially is due to a three-dimensional effect not captured in our two-dimensional model. Nonetheless, the strength of radial diffusion is largely independent of scale both when simulating two and three dimensions.

At a radius of~$30~\au$ in the simulation performed by \citetalias{Flock2020}, a giant vortex is induced by the Rossby wave instability. Dust diffusion in this vortex is independent of both direction and scale, with the radial diffusion being stronger at this radius than at~\mbox{$r=55~\au$} where its source is the vertical shear instability. Furthermore, the vertical diffusion coefficients evince that the dust layer possesses a Gaussian shape centred on the disk mid-plane rather than the typical wave-like shape caused by the vertical shear instability.

\section{Discussion}
\label{sect:discussion}
\subsection{Implications for modelling gas turbulence and dust diffusion in protoplanetary disks}
Quantitative analyses of protoplanetary disk turbulence are most commonly based on the~$\alpha$-model formulated by \citet{Shakura1973}. This model was devised to describe turbulence as an effective viscosity that facilitates stellar accretion by enabling angular momentum transport away from the star. Turbulence does not necessarily entail outward angular momentum transport, though -- while the turbulence induced by the vertical shear instability indeed does \citep{Nelson2013, Stoll2014}, the streaming instability on average causes gas to move outwards, not inwards \citepalias{Schafer2020}.

Furthermore, the $\alpha$-model is not designed to capture anisotropy and scale-dependence. Accurately describing, for instance, the strength of anisotropic turbulence induced by the vertical shear instability (\citealt{Nelson2013}, \citealt{Stoll2017}, \citetalias{Schafer2020}) thus requires reporting at least an~$\alpha$-parameter each for the radial and for the vertical dimension. We note that, moreover, two different definitions of the~$\alpha$-parameter exist. In studies of the vertical shear instability, it is most commonly defined based on the Reynolds stress to quantify angular momentum transport \citep[e.g.][]{Nelson2013, Stoll2014, Stoll2016, Stoll2017}. Yet, it can also be defined as the square of the Mach number when considering a mixing length approach and assuming the eddy turnover timescale to be similar to the inverse of the orbital frequency.

Similar to measuring gas turbulence, translating from its strength to the strength of the dust diffusion it entails is challenging. Taking our model of the streaming instability as an example, both gas turbulence and diffusion of millimetre-sized dust are largely isotropic, in agreement with previous models (\citealt{Johansen2007a}, \citealt{Schreiber2018}, \citetalias{Schafer2020}). In contrast, the centimetre-sized is diffused more strongly in the radial than in the vertical dimension \citep[see also][]{Schreiber2018, Li2021}. In addition to vertical diffusion being counteracted by dust sedimentation, \citet{Li2021} speculate that dust accumulation in clumps -- which is strong for the combination of dust-to-gas surface density ratio and dust size in this case \citepalias{Schafer2022} -- affects both the radial and the vertical diffusion coefficient. On the one hand, the radial coefficient is enhanced since dust particles embedded in clumps radially drift more slowly than isolated particles. On the other hand, the vertical coefficient is reduced because clumps are concentrated in the disk mid-plane.

Finally, our study highlights that the shape of a dust layer in the vertical dimension can not necessarily be expressed using a single scale height. The scale height relative to the mid-plane can be substantially greater than the one relative to the mid-plane of the dust layer, most notably by more than order of magnitude in our simulation of the vertical shear instability including centimetre-sized dust. This instability gives rise to strong large-scale dust diffusion resulting in a wave-shaped layer, but comparatively weak small-scale diffusion. Nonetheless, the streaming instability causes similarly strong small-scale and large-scale diffusion, resulting in the two scale heights being comparable.

Dust diffusion has been argued to impede the formation of planetesimals. Linear analyses including an $\alpha$-model prescription of turbulence reveal reduced growth rates of the streaming instability \citep{Umurhan2020, Chen2020}, and indeed planetesimal formation has been demonstrated to be inhibited by driven Kolmogorov-like turbulence in numerical simulations of the non-linear instability \citep{Gole2020}. \citet{Gerbig2020} and \citet{Klahr2020} derive criteria for gravitational collapse that expand on the Roche criterion by requiring dust self-gravity to overcome not only the stellar tidal forces but also diffusion.

Nevertheless, instabilities like the streaming instability, the vertical shear instability, and the convective overstability are not only sources of diffusion, but also drive planetesimal formation either directly or by causing pressure bumps and vortices in which dust accumulates (\citetalias{Schafer2020}, \citealt{Raettig2021}, \citealt{Lehmann2022}, \citetalias{Schafer2022}). Indeed, in \citetalias{Schafer2022} we find that dust concentration is stronger and planetesimal formation therefore possible for smaller dust sizes and lower dust-to-gas surface density ratios when simulating the vertical shear instability and the streaming instability in concert than when modelling the streaming instability alone. This is despite this work showing that diffusion is stronger when both instabilities are active.

\subsection{Limitations}
The most significant limitation of our study is the assumption of axisymmetry. While this assumption is justified when considering the linear regimes of the vertical shear instability \citep{Nelson2013, Barker2015} and the streaming instability \citep{Youdin2005}, the non-linear regimes exhibit deviations from this symmetry \citep{Nelson2013, Stoll2014, Kowalik2013}. We address this shortcoming by comparing our two-dimensional model of the vertical shear instability to the three-dimensional one presented by \citetalias{Flock2020}. This comparison shows that our model reproduces the strength of vertical dust diffusion and the resulting dust scale heights, though not the strength of radial dust diffusion. Moreover, our model can not replicate the formation of vortices owing to the vertical shear instability (\citealt{Richard2016}, \citealt{Latter2018}, \citetalias{Flock2020}), with these vortices affecting dust diffusion as noted in Sect.~\ref{sect:comparison}. 

For simplicity, we employ an isothermal equation of state when simulating the vertical shear instability (and an adiabatic equation of state to quench it when simulating the streaming instability in isolation). As detailed above, the more sophisticated radiation hydrodynamic model applied by \citetalias{Flock2020}, which provides less ideal conditions for the vertical shear instability to develop, nonetheless yields similar results. However, even this model does not include the coupling between dust diffusion owing to the vertical shear instability and gas cooling via gas-dust collisions regulating the growth of the instability \citep{Fukuhara2021, Fukuhara2024, Pfeil2023, Pfeil2024}. Self-consistent models by \citet{Fukuhara2024} and \citet{Pfeil2024} indeed demonstrate that, for millimetre-sized and centimetre-sized as we consider in this work which is only weakly coupled to the gas, the vertical shear instability possibly can not grow rapidly enough and induce sufficient diffusion to prevent run-away dust settling.

Further limitations of our study include not modelling other sources of gas turbulence and dust diffusion in protoplanetary disks besides the streaming instability and the vertical shear instability as well as considering only uniform dust sizes and neglecting dust evolution. Previous studies have explored the interplay between the magnetorotational instability and the vertical shear instability \citep{Cui2021, Cui2022}, between the magnetorotational instability and the streaming instability \citep{Johansen2007b, Johansen2011, Balsara2009, Tilley2010, Yang2018}, as well as between the convective overstability and the streaming instability \citep{Raettig2021}. Dust size distributions have been included in models of both the streaming instability \citep[e.g.,][]{Bai2010b, Schaffer2018, Schaffer2021, Krapp2019, Zhu2021, Yang2021} and the vertical shear instability \citep[e.g.,][]{Stoll2016, Flock2017, Flock2020}, with \citet{Pfeil2023, Pfeil2024} additionally simulating dust growth. 

\section{Conclusion}
\label{sect:conclusion}
Building on our previous work \citep{Schafer2020, Schafer2022}, in this study we analyse gas turbulence and dust diffusion in two-dimensional, axisymmetric global models of the vertical shear instability and the streaming instability with a dust-to-gas surface density ratio of~$2\%$. The focus of our study lies on diffusion and its the dependence on both direction and scale.

The vertical shear instability, on the one hand, is a source of anisotropic turbulence \citep{Nelson2013, Stoll2017, Schafer2020} and dust diffusion, both are stronger in the vertical than in the radial dimension. A central finding of our study is that, while vigorous large-scale vertical diffusion leads to the dust layer possessing the shape of a wave \citep{Flock2017, Flock2020, Dullemond2022}, the instability induces considerably weaker diffusion on small scales in the vertical dimension as well as in the radial dimension. We measure dimensionless diffusion coefficients of the order of~$10^{-3}$ when considering the large-scale vertical diffusion, but otherwise coefficients of~\mbox{${\sim}5\times10^{-4}$} in a simulation of millimetre-sized dust and vanishingly low ones in a simulation of centimetre-sized dust. Therefore, the dust layer is too thin to be resolved in the latter simulation despite the application of mesh refinement based on dust concentration.

The streaming instability, on the other hand, diffuses dust in a scale-independent manner, with the dust layer thus exhibiting a Gaussian shape \citep{Bai2010b, Li2021, Schafer2022}. Both turbulence \citep{Johansen2007a, Schafer2020, Yang2021, Baronett2024} and diffusion of millimetre-sized dust are further independent of direction, while centimetre-sized dust is diffused more strongly in the radial direction than in the vertical one \citep[see also][]{Schreiber2018, Li2021, Yang2021, Gerbig2023, Baronett2024}. The dimensionless diffusion coefficients amount to~\mbox{${\sim}10^{-5}$} for the smaller dust size, and to~\mbox{$5\times10^{-5}$} in the radial dimension and to~\mbox{$4\times10^{-4}$} in the vertical dimension for centimetre-sized dust.

We further consider models in which both instabilities are active. As we show in \citet{Schafer2020}, the streaming instability is the dominant source of turbulence in the dust layer in the scenario \emph{SIwhileVSI}, but the vertical shear instability in the scenario \emph{SIafterVSI}. While dust diffusion is thus overall stronger in the latter scenario, both scenarios exhibit contributions of both instabilities to this diffusion. Most notably, the streaming instability compensates for the lack of small-scale diffusion owing to the vertical shear instability. This way, the undulating mid-plane layer internally acquires a width that is set by the diffusion caused by the streaming instability. Whether planetesimals can form in such an undulating layer of high density will require future computer simulations that include both large vertical extents (for the vertical shear instability to thrive), good resolution of the layer width (for the streaming instability) and the inclusion of the self-gravity of the dust particle component.

\begin{acknowledgements}
We thank the anonymous referee for their constructive feedback that helped improve this paper. To analyse and visualise the simulations presented in this paper, the Python packages yt\footnote{\url{http://yt-project.org}} \citep{Turk2011}, Matplotlib\footnote{\url{https://matplotlib.org}} \citep{Hunter2007}, and NumPy\footnote{\url{https://numpy.org}} \citep{Oliphant2006} have been used. The FLASH Code has in part been developed by the DOE NNSA-ASC OASCR Flash Center at the University of Chicago. Computational resources employed to conduct the simulations were provided by the Regionales Rechenzentrum at the University of Hamburg, by the Norddeutscher Verbund für Hoch- und H\"ochstleistungsrechnen (HLRN), and by the SCIENCE HPC Center at the University of Copenhagen. U.S.~and A.J.~are thankful for funding from the Danish National Research Foundation (DNRF Chair Grant DNRF159). A.J.~further gratefully acknowledges funding from the Knut and Alice Wallenberg Foundation (Wallenberg Scholar Grant 2019.0442), the G\"oran Gustafsson Foundation, and the Carlsberg Foundation (Semper Ardens: Advance grant FIRSTATMO). M.F.~acknowledges support from the European Research Council (ERC), under the European Union’s Horizon 2020
research and innovation program (grant agreement No. 757957).

\end{acknowledgements}

\bibliography{references}

\begin{appendix}
\onecolumn
\section{Gas kinetic energy power spectra}
\label{sect:power_spectra}
To further examine the direction- and scale-dependence of the turbulence induced by the streaming instability and the vertical shear instability in isolation and in conjunction, we compute power spectra of the gas kinetic energy. We note that these spectra are limited in their expressiveness because, firstly, the domains of the simulations we conducted are non-periodic and, secondly, the grid resolution is variable because of the application of mesh refinement. To remedy the latter, we superimpose a covering grid with the base resolution on the domain, interpolating to this resolution where necessary.

\begin{figure*}[h!]
  \begin{minipage}{0.49\textwidth}
    \centering
    \includegraphics[width=\textwidth]{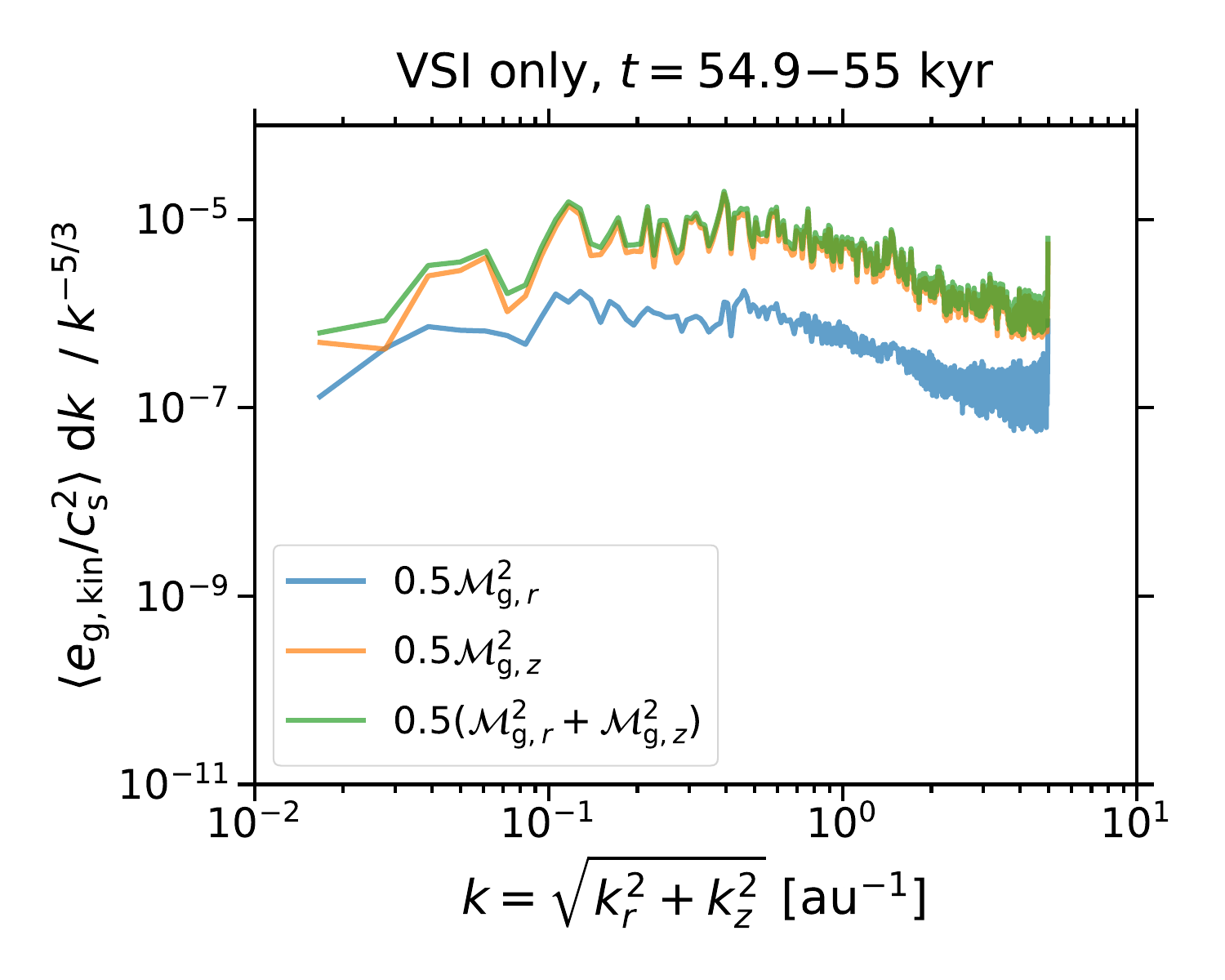} 
    \includegraphics[width=\textwidth]{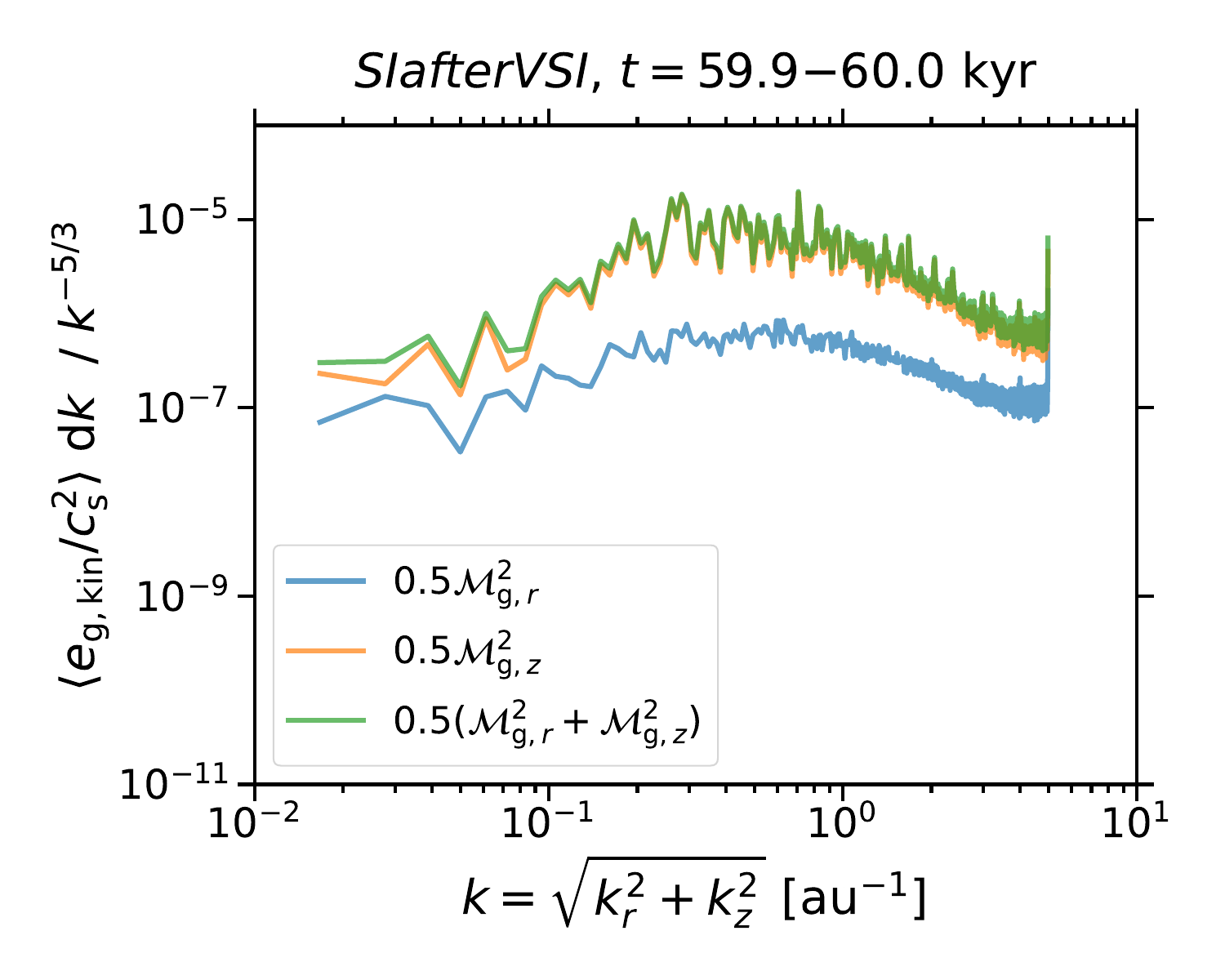}
  \end{minipage} 
  \hfill
  \begin{minipage}{0.49\textwidth}
    \centering
    \includegraphics[width=\textwidth]{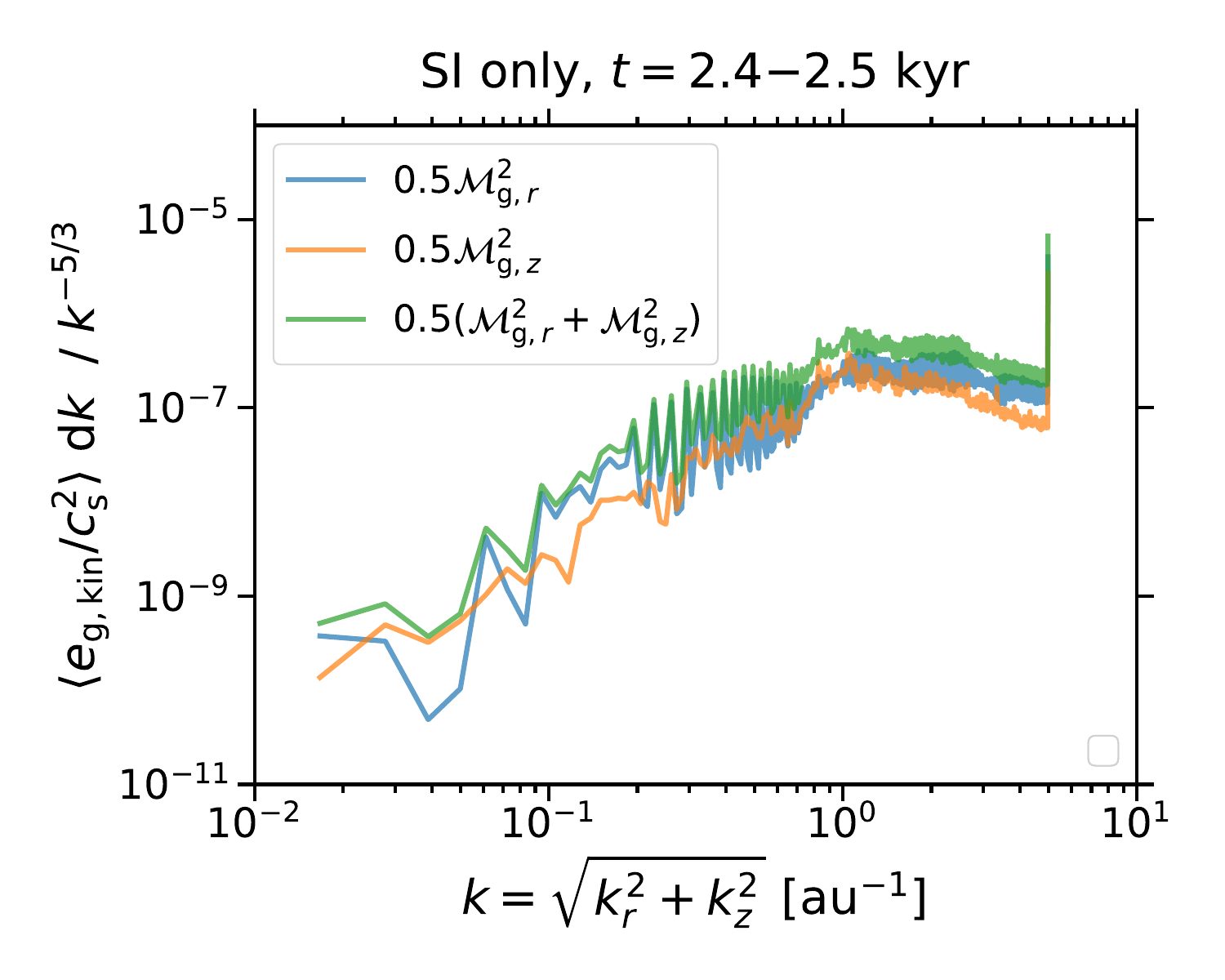} 
    \includegraphics[width=\textwidth]{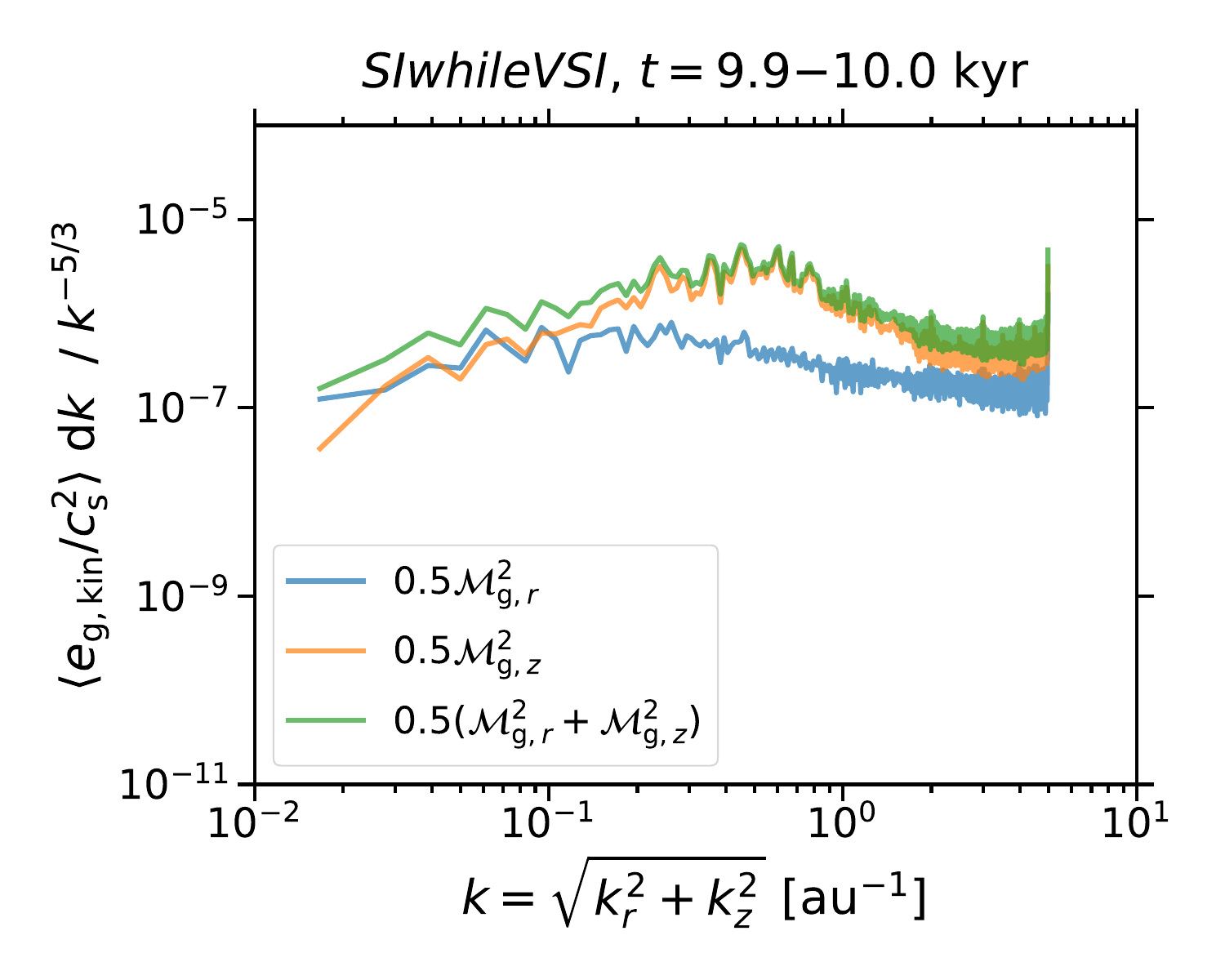}
  \end{minipage} 
  \caption{Power spectra of the gas specific kinetic energy~$e_{\textrm{g,kin}}$, non-dimensionalised using the square of the sound speed~$c_{\textrm{s}}$, as functions of the wavenumber~\mbox{$k=\sqrt{k_r^2+k_z^2}$}. The top left, top right, bottom left, and bottom right panels, respectively, show simulations of the vertical shear instability and of the streaming instability as well as of the scenarios \emph{SIafterVSI} and \emph{SIwhileVSI} with a dust size of~$3~\cm$. The kinetic energy is calculated taking into account different velocity components, with the resulting spectra being plotted using differently coloured lines. All spectra are averaged over~$100~\yr$ and normalised to the spectrum of Kolmogorov turbulence~\mbox{$E_{\textrm{g,kin}}\propto k^{-5/3}$}. The spectra are similar in our simulations of only the vertical shear instability, of the scenario \emph{SIafterVSI}, and of the scenario \emph{SIwhileVSI} when considering scales greater than~\mbox{${\sim}1~\au$}, with the kinetic energy associated with the vertical velocity being higher by as much as an order of magnitude than the energy associated with the radial velocity. In comparison, the energies in the simulation of the streaming instability in isolation are largely equal for the two velocity components but generally lower; only on scales of~\mbox{${\sim}1~\au$} and smaller are they comparable to the energy in the radial motions owing to the vertical shear instability. At the same scales, the power spectra in the simulation of the scenario \emph{SIwhileVSI} seem a mixture of the spectra in the simulations of the streaming instability and the vertical shear instability individually in terms of magnitude and direction-dependence of the kinetic energy.}
  \label{fig:power_spectra}
\end{figure*}

In Figure~\ref{fig:power_spectra}, we depict the kinetic energy power spectra obtained at the end of our simulations of only the vertical shear instability and the streaming instability as well as of the scenarios \emph{SIwhileVSI} and \emph{SIafterVSI}, all including dust with a size of~$3~\cm$. The spectra are normalised to the spectrum of Kolmogorov turbulence~\mbox{$e_{\textrm{g,kin}}\propto k^{-5/3}$}, where~$e_{\textrm{g,kin}}$ is the gas specific kinetic energy and the wavenumber~\mbox{$k=\sqrt{k_r^2+k_z^2}$}. This spectrum is characteristic of freely evolving three-dimensional turbulence, though, while two-dimensional turbulence exhibits an inverse energy cascade~\mbox{$e_{\textrm{g,kin}}\propto k^{-5/3}$} from the injection scale towards larger scales and an enstrophy cascade -- enstrophy can be defined as the square of the vorticity -- with~\mbox{$e_{\textrm{g,kin}}\propto k^{-3}$} towards smaller scales \citep{Kraichnan1967, Lyra2019}.

Consistent with previous work (\citealt{Nelson2013}, \citealt{Stoll2017}, \citetalias{Schafer2020}) and with what we discuss in Sect.~\ref{sect:gas}, the vertical shear instability gives rise to anisotropic turbulence, with the kinetic energy associated with vertical motions being greater by up to an order of magnitude than that associated with radial motions on all scales. In comparison, the turbulence induced by the streaming instability is both weaker \citepalias{Schafer2020} and mostly isotropic (\citealt{Johansen2007a}, \citetalias{Schafer2020}, \citealt{Schreiber2018}, \citealt{Yang2021}, \citealt{Baronett2024}). Only on scales of~$1~\au$ and smaller are the kinetic energies in the simulation of the streaming instability comparable to the energy in the radial motions in the simulation of the vertical shear instability.

These findings are in line with the vertical shear instability diffusing dust more strongly on large vertical scales than the streaming instability (see Sect.~\ref{sect:vertical_diffusion}). On the other hand, they render the inefficacy of the vertical shear instability to diffuse centimetre-sized dust in the radial dimension (see Sect.~\ref{sect:radial_diffusion}) and on small scales in the vertical dimension even more striking. 

The power spectra in the scenario \emph{SIafterVSI} resemble those in the simulation of the vertical shear instability in isolation. That is, they do not reflect that, while the turbulence elsewhere is predominantly caused by the vertical shear instability, the streaming instability is the main source of turbulence locally in dust overdensities \citepalias{Schafer2020}. This is likely because the contribution of the kinetic energy in these overdensities to the power spectra is negligible, particularly since we apply a covering grid with the base resolution to obtain the spectra.

On scales larger than~$1~\au$, the turbulence in the scenario \emph{SIwhileVSI} as well exhibits power spectra comparable to the ones of the turbulence induced by the vertical shear instability. On smaller scales, though, the spectra appear to be a superposition of those in our simulation of this instability and in our simulation of the streaming instability. Indeed, we find the turbulence to be driven by the vertical shear instability away from the dust layer in this scenario, but mostly by the streaming instability in this layer \citepalias{Schafer2020}.
\end{appendix}
\end{document}